\documentclass[twoside,leqno,twocolumn]{article}

\usepackage[letterpaper]{geometry}

\usepackage{siamproceedings}

\usepackage[T1]{fontenc}
\usepackage{amsfonts}
\usepackage{epstopdf}

\usepackage{multirow}
\usepackage{amsmath,amsfonts}
\usepackage[endLComment=]{algpseudocodex}

\patchcmd{\subsubsection}{\itshape}{\bfseries}{}{}
\usepackage{graphicx}
\usepackage{textcomp}
\usepackage{xcolor}
\usepackage{comment}
\usepackage{soul}
\usepackage[inline]{enumitem}
\usepackage{hyperref}
\usepackage{cleveref}[2012/02/15]
\crefformat{footnote}{#2\footnotemark[#1]#3}
\usepackage{footnote}
\makesavenoteenv{tabular}
\makesavenoteenv{table}
\usepackage{booktabs}
\usepackage{adjustbox}
\newcommand{\AlgIn}{\Statex \textbf{Input:} }
\newcommand{\AlgOut}{\Statex \textbf{Output:} }
\DeclareMathOperator*{\argmax}{arg\,max}

\newcommand{\calN}{\mathcal{N}}

\newcommand\todo[1]{\textcolor{red}{#1}}

\newcommand{\linkfoot}[1]{\footnote{\href{#1}{#1}}}

\usepackage{latexsym}
\usepackage{verbatim}
\usepackage{balance}

\usepackage{multicol}

\usepackage[c3, nocomma, short]{optidef}

\usepackage{xspace}
\usepackage{enumitem}
\usepackage[font={small}]{caption}

\newcommand{\apj}{\textsc{ApproxJoin}}
\newcommand{\tokenjoin}{\textsc{TokenJoin}}

\newcommand{\ajgd}{\textsf{AJ--GD}\xspace}
\newcommand{\ajld}{\textsf{AJ--LD}\xspace}
\newcommand{\ajps}{\textsf{AJ--PS}\xspace}
\newcommand{\tjhg}{\textsf{TJPJ--HG}\xspace}
\newcommand{\tjev}{\textsf{TJPJ--EV}\xspace}
\newcommand{\tjpj}{\textsf{TJPJ}\xspace}

\theoremstyle{plain}
\newtheorem{observation}{Observation}
\usepackage[compact]{titlesec}
\titlespacing{\section}{0pt}{2ex}{1ex}
\titlespacing{\subsection}{0pt}{2ex}{1ex}
\titlespacing{\subsubsection}{0pt}{1ex}{1ex}

\ifpdf
  \DeclareGraphicsExtensions{.eps,.pdf,.png,.jpg}
\else
  \DeclareGraphicsExtensions{.eps}
\fi

\newsiamremark{remark}{Remark}
\newsiamremark{hypothesis}{Hypothesis}
\crefname{hypothesis}{Hypothesis}{Hypotheses}
\newsiamthm{claim}{Claim}

\usepackage{amsopn}

\begin{document}

\newcommand\relatedversion{}
\renewcommand\relatedversion{\thanks{The full version of the paper can be accessed at \protect\url{https://arxiv.org/abs/0000.00000}}} 

\title{\Large \textsc{ApproxJoin}: Approximate Matching for\\Efficient Verification in Fuzzy Set Similarity Join}
    \author{Michael Mandulak\thanks{Rensselaer Polytechnic Institute\\(\email{mandum@rpi.edu}, \email{slotag@rpi.edu}).}
    \and S M Ferdous\thanks{Pacific Northwest National Laboratory\\
  (\email{sm.ferdous@pnnl.gov}, \email{sayan.ghosh@pnnl.gov},
  \email{hala@pnnl.gov}).}    
    \and Sayan Ghosh\footnotemark[2]  
    \and Mahantesh Halappanavar\footnotemark[2]  
    \and George Slota\footnotemark[1]}

\date{}

\maketitle 







\begin{abstract}

The set similarity join problem is a fundamental problem in data processing and discovery, relying on exact similarity measures between sets. In the presence of alterations, such as misspellings on string data, the fuzzy set similarity join problem instead approximately matches pairs of elements based on the maximum weighted matching of the bipartite graph representation of sets. State-of-the-art methods within this domain improve performance through efficient filtering methods within the filter-verify framework, primarily to offset high verification costs induced by the usage of the Hungarian algorithm - an optimal matching method. Instead, we directly target the verification process to assess the efficacy of more efficient matching methods within candidate pair pruning. 

We present \apj, the first work of its kind in applying approximate maximum weight matching algorithms for computationally expensive fuzzy set similarity join verification. We comprehensively test the performance of three approximate matching methods: the Greedy, Locally Dominant and Paz Schwartzman methods, and compare with the state-of-the-art approach using exact matching. Our experimental results show that \apj~yields performance improvements of 2-19$\times$ the state-of-the-art with high accuracy (99\% recall). 

\end{abstract}



\section{Introduction.}
Given two collections of sets, {\em set similarity join} computes all pairs of similar sets, where two sets are considered similar if the overlap of set elements (or tokens) is above a user-defined threshold using a user-defined metric such as Jaccard similarity (detailed in \S\ref{ssec:set_sim_prelim}). 
Set similarity join is a foundational database operator used widely in data science, data mining, and data management for tasks such as data cleaning and entity resolution \cite{Chaudhuri2006, Sarawagi2004}.
Most set similarity join approaches follow the {\em filter-verify} framework~\cite{Mann2016}, where the filtering phase (e.g., prefix-filters) generates (possibly smaller) candidate pair sets, which are then verified (e.g., similarity computation using Jaccard function) to be included in the solution set if the similarity threshold is met. Traditional approaches rely on the notion of exact similarity and direct equivalence of elements with the sets. In the real world, however, exact similarity may not be robust in the presence of noise such as misspellings, format variations, and other types of data alterations. Fuzzy set similarity joins, which are based on Fuzzy token matching \cite{Wang2011},  have been proposed to tackle these issues. 

The most popular Fuzzy token matching approaches rely on the {\em Bipartite matching}-based similarity scores \cite{FastJoin,SilkMoth,MF-Join,TokenJoinOG,melnik2002similarity,sarma2012finding}, where a matching $M$ is a subset of edges in a bipartite graph such that no two edges in $M$ are incident on the same vertex.
Here, we first compute a bipartite graph from the two sets (which are part of the collection of sets; an example is shown in \S\ref{fig:bipartite-matching-example}) using their elements (or tokens) as the vertices in the two vertex partitions of the graph. The edge weights between each pair of vertices are computed relative to a chosen similarity measure between the pairs of elements. A maximum weight bipartite matching (defined in \S\ref{ssec:bipartiteMatching}) is computed on this graph, and the weight of the matching is incorporated into the similarity score. However, unlike traditional approaches, matching based similarity is compute and resource intensive. For example, given two sets each having $n$ elements, the Jaccard index between them can be computed in $O(n \log n)$ time, by sorting them both in lexicographical order and then merging them to find the union and intersection. In contrast, the primal-dual Hungarian algorithm for bipartite weighted matching (on a complete bipartite graph) requires $O(n^3)$ time. Similarly, from a space complexity point of view, the Jaccard based approach requires $O(n)$ space, whereas the optimal matching needs to construct the complete graph, requiring $O(n^2)$ space (\S\ref{sec:methods}).

\begin{figure}[!ht]
    \centering
     \hspace{0.8cm} \includegraphics[width=0.8\linewidth]{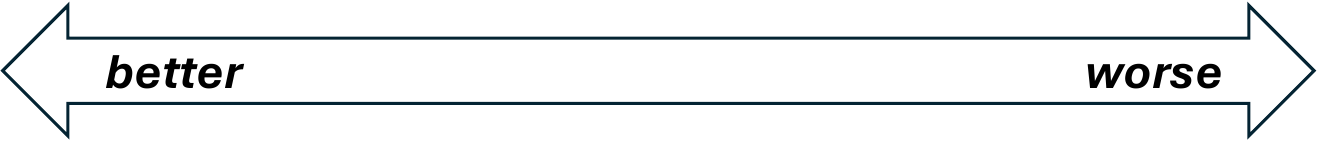}
    \includegraphics[width=0.9\linewidth]{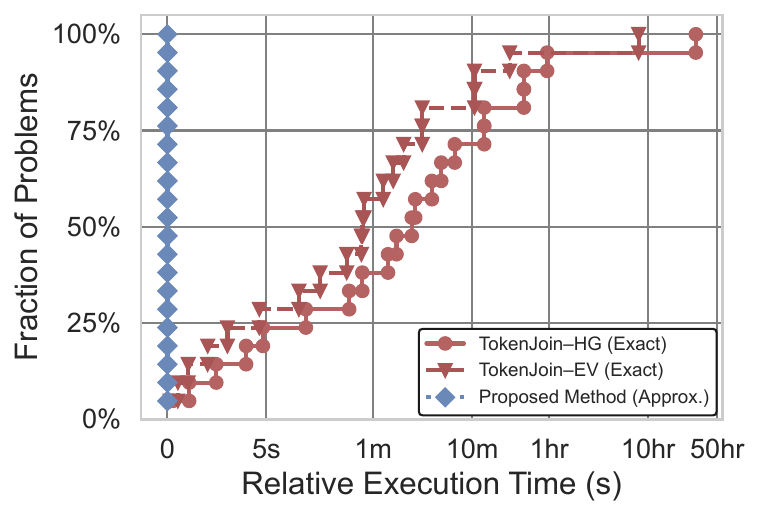}
    \caption{Approximate methods vastly outperform exact methods for fuzzy set similarity join. 
Performance profile~\cite{performanceprofiles} summarizes impact of proposed \apj~ vs. state-of-the-art \textsc{TokenJoin} methods~\cite{TokenJoinOG}, details in \S\ref{sssec:perf_summary}.}
    \label{fig:introRuntimePerf}
\end{figure}

The high complexity of the matching-based approaches motivates earlier work to design better and more aggressive filtering techniques for set similarity joins, where the primary goal is to avoid the matching computations (in the verification phase) whenever possible. In this paper, 
we directly address this fundamental limitation by improving the cost of computing weighted matching instead of improving filtering approaches (detailed in \S\ref{sec:methods}) further. We engineer the algorithmic pipeline by employing {\em approximate} weighted matching and demonstrate that the runtime can be dramatically decreased, e.g. from \textbf{35 hours} to \textbf{5 hours} for the KOSARAK dataset(detailed in \S\ref{sec:evals}) without affecting the accuracy. As depicted in Fig.~\ref{fig:introRuntimePerf}, we observe consistently better performance for the proposed methods relative to the state-of-the-art methods. We implement three representative approximate matching algorithms for computing set similarity and evaluate their performance within the state-of-the-art \tokenjoin~\cite{TokenJoinOG} framework (detailed \S\ref{sec:methods-matching}). Using a large set of input datasets, we demonstrate the effectiveness in terms of the quality, execution time, and memory usage of the approximate matching-based set similarity join, which we call \apj{} (detailed \S\ref{sec:evals}).

To the best of our knowledge, this is the first work of its kind to apply approximate matching algorithms to fuzzy set similarity joins. The main contributions of this work are:
\begin{itemize}[leftmargin=*,topsep=2pt]
    \item We propose \apj{}, an extension of the verification process within state-of-the-art fuzzy set similarity join using approximate matching methods.
    \item We implement the Locally-Dominant (LD), Paz Schwartzman (PS) and Greedy (GD) approximate matching methods to compare execution times with the \tokenjoin~ defaults of the Hungarian and Efficient Verification methods.
    \item We demonstrate a range of \textbf{2-19}$\times$performance improvements in verification and total execution times compared to the optimal methods on a variety of sparse and dense inputs using our publicly available implementation\footnote{\href{https://github.com/mmandulak1/sasimi}{https://github.com/mmandulak1/sasimi}}. We also compare the qualities of approximate methods relative to the optimal, showing average recall values above \textbf{0.99}.
\end{itemize}
\section{Preliminaries. }\label{background}


\subsection{Fuzzy Set Similarity Join.}
\label{ssec:set_sim_prelim}

\begin{figure}[!ht]
    \centering
    \includegraphics[width=\linewidth]{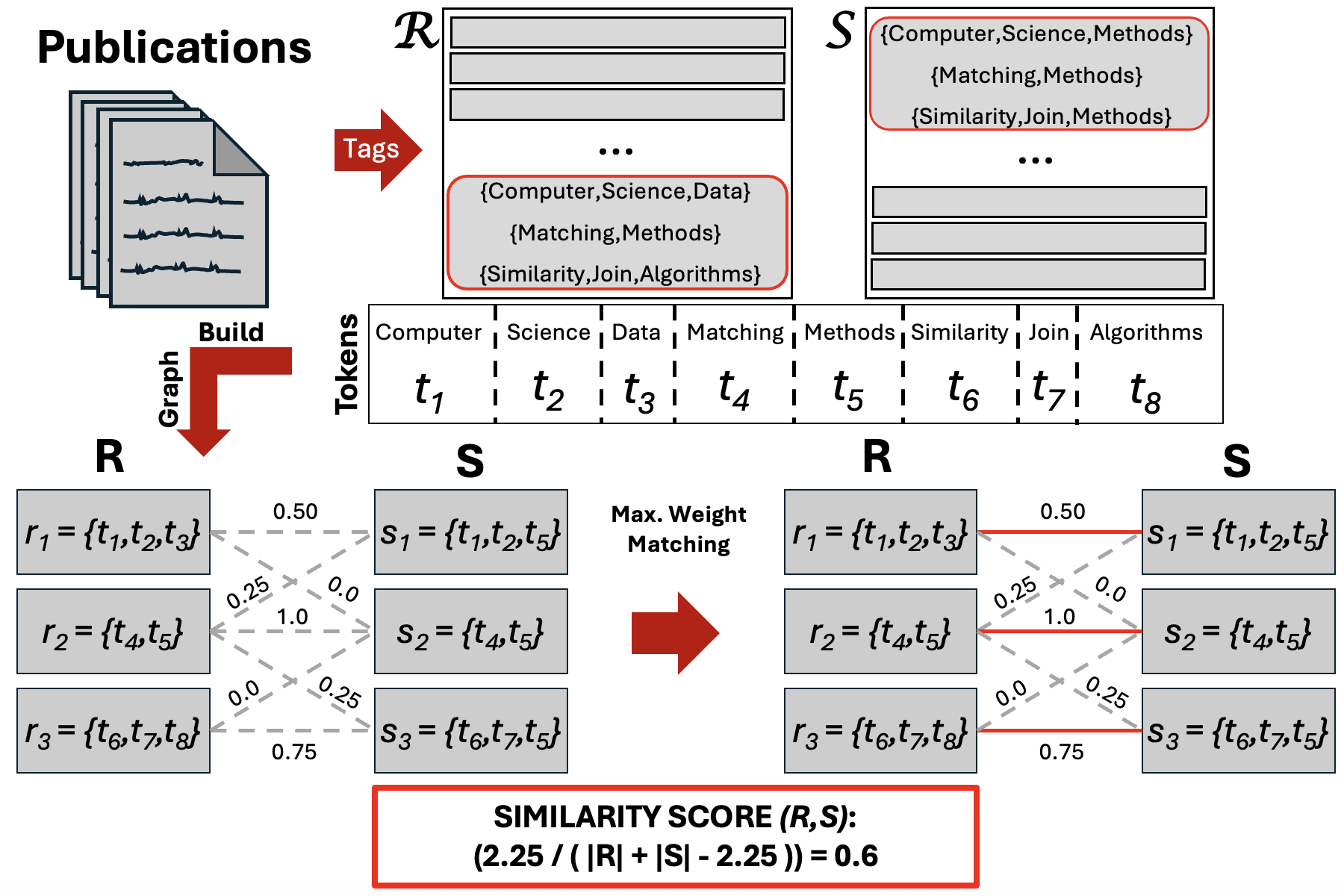}
    \caption{Illustration depicting the usage of matching within the fuzzy set similarity join workflow. Text data, such as publication tags, are split into tokens and further into a bipartite graph with similarity-based weights. The maximum weight matching is incorporated into the resultant fuzzy similarity score between the sets $R$ and $S$.}
    \label{fig:bipartite-matching-example}
\end{figure}

Our inputs for Fuzzy set similarity join problem are two collections of sets ($\mathcal{R}$ and $\mathcal{S}$), a set similarity function $sim_{\phi}(R,S)$ (where $R\in \mathcal{R}$ and $S\in \mathcal{S}$), and a user-defined threshold $\delta$. Our goal is to compute all pairs of similar sets, which is defined as:
$\mathcal{R} \Join_{\delta} \mathcal{S} = \{(R, S) \in \mathcal{R} \times \mathcal{S} ~ \vert ~sim_{\phi}(R,S) \geq \delta\}$ \cite{TokenJoinOG,SilkMoth,relatedness_definition}.
A set $R \in \mathcal{R}$ contains elements $r \in R$, where each element can be composed of a set of tokens $t \in r$ that are designated as $q$-grams in string tokenization. 
In this work, we focus on the {\em fuzzy set similarity problem} to approximately match set elements to each other, rather than the traditional definition that focuses on exact set overlap to compute $sim_{\phi}(R,S)$. We describe this process in detail as follows, but also provide a visual example in Fig.~\ref{fig:bipartite-matching-example}.

To approximately relate two sets $R$ and $S$ based on a threshold $\delta$, we can formulate a bipartite graph $G(V,E, w)$, $V= R \cup S$, between the elements of $R$ and $S$ with edges weighted by a chosen similarity measure $\phi(r,s) \in[0,1]$. There exist many measures for set similarity, but we focus on Jaccard similarity (JAC) and normalized edit similarity (NEDS), which are defined as follows relative to two elements $r,s$:
\begin{equation*}
    JAC(r,s) = \frac{|r \cap s|}{|r \cup s|} \; \text{,} \; NEDS(r,s) = 1-\frac{LD(r,s)}{\text{max}(|r|,|s|)}.
\end{equation*}
For NEDS, $LD$ is the Levenshtein distance \cite{Levenshtein_distance}, which is the number of \textit{edit} operations (delete, insert, substitute) to transform one string into the other.

After constructing the bipartite graph $G$, we find a maximum weight bipartite matching $M$ on $G$ to compute a similarity score between the pair $(R,S)$ (defined in \S\ref{ssec:bipartiteMatching}). The total weight of the matching, $|R~\widetilde{\cap}_{\phi}~S|$, is incorporated into set similarity between $R$ and $S$ as follows:
\begin{equation*}
    sim_{\phi}(R,S) = \frac{|R~\widetilde{\cap}_{\phi}~S|}{|R| + |S| - |R~\widetilde{\cap}_{\phi}~S|}.
\end{equation*}
Using this notion of fuzzy set similarity, we can define the problem of fuzzy set similarity join as follows:
\newtheorem{problem}{Problem}\label{problem:tokenjoin}
\begin{problem}{(Threshold-Based Fuzzy Set Similarity Join)}
\\Given two collections of sets $\mathcal{R}$ and $\mathcal{S}$, a similarity function $sim_{\phi}(R,S)$, and a similarity threshold $\delta \in [0,1]$, find all pairs of sets: 
$\mathcal{R} \widehat{\Join}_{\delta} \mathcal{S}$ $=$ $\{(R, S) \in \mathcal{R} \times \mathcal{S} ~ \vert ~sim_{\phi}(R,S) \geq \delta\}$.
\end{problem}

\begin{table}[!h]
  \centering
  \begin{tabular}{c c}
  \toprule
    \textbf{Notation}  & \textbf{Description} \\
    \midrule
    $\mathcal{R}$, $\mathcal{S}$ & Collections of sets \\
    $R\in \mathcal{R}$, $S\in \mathcal{S}$ & Sets within a collection \\
    $r\in R$, $s\in S$ & Elements within a set \\

    \bottomrule
  \end{tabular}
   \caption{Key notations used in the paper.}
  \label{tab:notation}
  \vspace{-1em}
\end{table}

\subsection{Bipartite Weighted Matching.}
\label{ssec:bipartiteMatching}
A bipartite graph $G(V,E, w)$ is defined on a vertex set $V= R \cup S$, where $R$ and $S$ are the two disjoint parts. The edge set $E$ consists of sets $\{u,v\}$, where $u \in R$ and $v \in S$. $w: E \to R_{\geq 0}$ is a non-negative weight function defined on the edges. Throughout the paper, we denote $n \coloneqq |V|$, and $m \coloneqq |E|$. For a vertex $v$, let $\delta(v)$ be the set of edges incident on $v$.

 A \emph{matching} $M$ in $G$ is a subset of $E$, where for each edge in $M$ is vertex disjoint, i.e, $e_i \cap e_j = \emptyset$, where $i\neq j$ and $e_i,e_j \in M$. A \emph{perfect matching} is a matching that covers all the vertices. Formally, if $M$ is perfect then $\bigcup \{e\in M\} = V$. Let $w(M)$ be the sum of weights of the edges in the matching, i.e., $w(M) = \sum_{e \in M} w(e)$. The maximum weight bipartite matching (MWBM) is to find a matching $M_*$, whose $w(M_*)$ is the maximum among all possible matchings of $G$. Note that the graphs generated from similarity joins are complete bipartite graphs, and the resultant maximum weighted matching is always perfect. Such problems are also referred to as the assignment problem in the literature. For a constant $0<\alpha<1$, an \emph{approximate weighted matching} $M_a$ is a matching whose weight is at least $\alpha$ fraction of maximum weight, i.e., $w(M_a) \geq \alpha \cdot w(M_*)$.
 We detail some of the algorithmic developments of matching in \S\ref{ssec:LP} alongside a brief overview of the primal-dual approach.

\section{Methodology.}
\label{sec:methods}

In this section, we discuss our primary contribution -- the \apj~(AJ) method. We discuss in \S\ref{sec:methods-threshold} the \tokenjoin~ methodology that we use for candidate generation and filtering, which are directly incorporated from \tokenjoin,~ followed by our proposed algorithms for approximate matching.

\subsection{Set Verification}\label{sec:methods-verification}
We outline the standard verification phase  in a matching-based fuzzy set similarity join in Algorithm \ref{alg:verify}. The goal of verification is to score two candidate sets, $R$ and $S$, based on approximate similarity. First, the algorithm proceeds with a deduplication procedure, which eliminates the duplicate elements between $R$ and $S$. In practice, each duplicate element yields a similarity measure of 1 to itself, which is incorporated into the overall similarity through the overlap calculation. Following deduplication, if the size of $R_d$ is zero (we guarantee prior to verification that the smallest set of the pairing is labeled as $R$), then there is no matching to be done, and we output the scoring based on the overlaps. Otherwise, we proceed to build the graph, where we assign a similarity score as edge weight between each element (the vertices) in the $R_d$ and $S_d$ records. We also maintain an upper bound (UB) on maximum weighted matching through the maximum score generated which allow us to skip matching if the UB is smaller than the threshold $\delta$ (as shown in \cite{TokenJoinOG}). Finally, we compute the matching (if not skipped) on the graph and include the number of overlaps to generate a score of overall similarity between $R$ and $S$. We provide a description of matching methods through the illustration in Fig.~\ref{fig:verifyCartoon}, but detail as follows.

\subsection{Matching Algorithms}\label{sec:methods-matching}
We formally define the matching problem in \S\ref{ssec:bipartiteMatching} and provide a brief literature review in \S\ref{sec:Matchings}. In this section, we detail the matching algorithms in the context of set similarity join. 
The state-of-the-art bipartite matching based fuzzy set join methods, such as \cite{TokenJoinOG,SilkMoth}, employ  a primal-dual based method \cite{OGBipartite} 
commonly referred to as the Hungarian (HG) algorithm. The HG algorithm is an iterative one, where each iteration requires $O(n^2)$ time with at most $n$ iterations giving the total $O(n^3)$ runtime. In this paper, we propose incorporating efficient approximate matching into the verification phase of the Fuzzy set similarity workflow. We choose three representative approximate maximum weight matching methods: Greedy (GD), Locally Dominant (LD) and the semi-streaming algorithm of the Paz and Schwartzman (PS). We briefly describe these matching algorithms alongside optimizations we made within the implementations.

\subsubsection{\textbf{Hungarian (HG) Method:}} 

The Hungarian (Kuhn-Munkres) \cite{kuhn1955hungarian} is a primal-dual algorithm for solving MWBM problem optimally. It starts with a set of feasible dual variables of the primal and dual formulations in Fig.~\ref{fig:match-formulations}. The primal solution $x$ (i.e., the matching $M$) is initialized with an empty set. We find a trivial feasible dual solution, which sets $y(u)$ for $u \in R$ as the maximum weight of its incident edges, and sets $y(v)=0$, for $v \in S$. Note that, we start with an infeasible primal but feasible dual solution. The goal is to go towards primal feasibility while maintaining the dual feasible solutions. To facilitate that, we maintain an equality graph $(G_\ell (V_\ell, E_\ell)$, which is a subgraph of $G$ containing the tight edges in terms of dual variables (i.e., the edges $e$ that satisfy $y(u) + y(v) = w(e)$). The HG algorithm runs in phases, where in each phase we have two possibilities: i) The matching (primal) is increased by one edge through finding an augmenting path in $G_{\ell}$, or ii) the equality subgraph is expanded to $G_{\ell'}$, where $E_\ell \subset E_{\ell'}$. The algorithm ends when $M$ is a perfect matching (without loss of generality, we assume that the matching here is maximum-weight perfect matching). Note that, during the algorithm we always maintain $w(M) = \sum_{v \in V} y(v)$. So, from Lemma~\ref{lem:duality}, when $M$ is a perfect matching (i.e., a primal feasible solution) the strong duality holds and we conclude that $M$ is also a maximum weight matching. It can be shown that there are at most $n$ phases and each phase can be implemented in $O(n^2)$ time resulting in the $O(n^3)$ overall runtime. We use the HG implementation provided in NetworkX \cite{Hagberg2008} for empirical evaluations.

\subsubsection{\textbf{Efficient Verification (EV) Method:}}
The authors of \tokenjoin~ \cite{TokenJoinOG} present an \textit{Efficient Verification} method based on the primal-dual basis of the Hungarian method. The EV method tailors the Hungarian method to the application, beginning with an empty matching and calculating a tentative verification score based on the current state of the matching. An upper bound for the score is generated by matching every unmatched vertex to its highest weight neighbor, allowing overlaps in the matching. A lower bound is similarly calculated by greedily matching all unmatched vertices. If the upper bound falls short of the threshold, then the candidate set pairing can be discarded. If the lower bound surpasses the threshold, then the candidate set pairing can be committed to the resultant set. This provides an early-termination condition for the traditional Hungarian approach. We specifically use the upper bound and lower bound (UBLB) variation, calculating both bounds and terminating the matching early if one of the two conditions are satisfied. We use the implementation provided by \cite{TokenJoinOG}.

\subsubsection{\textbf{Greedy (GD) Method:}}

The 1/2-approximate greedy matching algorithm \cite{avis1983survey} starts with an empty matching and follows a simple process: sort the edges of the graph in descending order by weight and add each non-conflicting edge into the matching in that order. In implementation, we designate a sorted list of edge tuples and simply iterate over the edges, checking if either of the endpoints are already matched. Finally, we return the weight of matching by summing the matching edge weights. 

\subsubsection{\textbf{Locally Dominant (LD) Method:}}
We describe the 1/2-approximate Locally Dominant (Pointer Chasing) method from \cite{PreisMatching} in Algorithm \ref{alg:ld}. This method focuses on finding locally dominant edges among neighborhood edges, where the algorithm proceeds in two phases: \textit{pointing} and \textit{matching}. In the pointing phase, each vertex scans through its neighborhood and sets a pointer to its highest weighted, unmatched neighbor (ties are broken consistently using vertex ids). In the matching phase, all mutually-pointing vertices are committed to the matching, and all non-mutual pointers are reset for the following iteration. In implementation, we build our graph from adjacency lists for neighborhood traversal each iteration. We substitute HG with LD in Line \ref{alglineV:matching} of Algorithm \ref{alg:verify}.

\begin{algorithm}[!ht]
    \caption{Verification with PS Matching}
    \label{alg:ps_verification}
    \begin{algorithmic}[1]
    \AlgIn{Candidate Sets: $R,S$, Threshold $\theta_R$, constant $\epsilon$}
    \AlgOut{Score: $sim_\phi(R,S)$}
        \State $R_d,S_d,\text{ov} = \texttt{deduplication}(R,S)$
        \State $ \text{UB} = |R|; M \gets \emptyset$; $\text{Stk} \gets \emptyset$; $\forall v \in V: \phi(v) = 0$
        \ForAll {$r \in R_d$}
        \label{alglinePSV:R_for_loop}
            \State $max_s = 0$
            \ForAll{$s \in S_d$}
                \State $w(e) = \texttt{sim}(r,s)$ \Comment{$e=\{r,s\}$}
                \State $max_s = \texttt{max}(max_s,w(e))$
                \If{$w(e)> (1+\epsilon)(\phi(r) + \phi(s))$} \Comment{Streaming Process}\label{alglinePSV:stream_process}
                    \State $w'(e) = w(e) - (\phi(r) + \phi(s))$
                    \State $\phi(r) = \phi(r) + w'(e);\phi(s) = \phi(s) + w'(e)$ \label{ln:update-dual}
                    \State \text{Stk}.push($e$) \label{alglinePSV:stack_commit}
                \EndIf
            \EndFor
            \State $\text{UB} = \text{UB} - (1-max_s)$
            \If{$\theta_R > \text{UB}$}
                \State \Return $\frac{\text{UB}}{(|R| + |S| - \text{UB})}$ \label{alglinePSV:ubReturn}
            \EndIf
        \EndFor
        \While{$\text{Stk} \ne \emptyset$} \Comment{Post Processing} \label{alglinePSV:post_process}
            \State $e(r,s) \gets \text{Stk.pop()}$
            \If{$(V(M) \cap \{r,s\}) = \emptyset$} \Comment{$V(M)$: vertex set covered by $M$. }
                \State $M \gets M \cup \{e\}$
            \EndIf
        \EndWhile
        \State \Return $\frac{w(M) + \text{ov}}{(|R| + |S| - (w(M)+\text{ov}))}$        
    \end{algorithmic}
\end{algorithm}

\subsubsection{\textbf{Semi-Streaming Paz and Schwartzman (PS) Method:}} 
\label{sec:ps}
Now, we turn to a streaming matching algorithm for bipartite weighted matching. In a (semi-) streaming matching setting, the edges are streamed one by one and the algorithm is required to compute the matching at the end of stream without storing the full graph into memory. Thus, the memory usage of this algorithm is required to be $O(n \log n)$. This model of computation is primarily motivated to solve extreme-scale graphs, where even storing the graph is infeasible. However, as detailed in~\cite{ferdousStreaming}, streaming algorithms could be an efficient alternative to offline algorithms.

We briefly describe the state-of-the-art $\frac{1}{2+\varepsilon}$-approximate semi-streaming algorithm due to Paz and Schartzman (PS)~\cite{PazS17}, and refer to~\cite{GhaffariW19,ferdousStreaming} for a more detailed explanation. In Alg.~\ref{alg:ps_verification}, we show a detailed pseudo-code of the PS algorithm integrated with the verification phase. The PS method begins with an empty stack, $Stk$, and initializes a set of variables, $\phi(.)$ (approximate dual values) to zero for each vertex of the graph. For an edge $(r,s)$ that arrives in the stream, we push the edge to $Stk$ if its weight is greater than $(1+\varepsilon)$ times $\left(\phi(r)+\phi(s)\right)$, otherwise we discard the edge. If $\{r,s\}$ survives, we update the $\phi(r)$ and $\phi(s)$ according to the line~\ref{ln:update-dual} as shown in Alg.~\ref{alg:ps_verification}. 

In the post-processing phase, we  compute a maximal matching in the stack order, by unwinding the stack one edge at a time and inserting into the matching if it does not violate the matching constraints. The PS algorithm requires linear time in the number of edges. The space efficiency comes from the fact that many edges are pruned during the streaming phase and only $O(\frac{n \log n}{\varepsilon})$ edges are survived in the stack. The $\varepsilon > 0$ here provides a trade-off between the memory efficiency and the approximation guarantee.

From the construction of the $\phi$ values, it can easily be shown that $(1+\varepsilon) \phi(v), 
\forall v\in V$ is a feasible dual of dual problem~\ref{form:dual-match}. So, using weak-duality of Lemma~\ref{lem:duality}, we derive the following observation:
\begin{observation}
\label{obs:dual-bound}
    The $(1+\varepsilon) \sum_{v \in V} \phi(v)$ is an upper bound on the maximum weight bipartite matching of the graph $G$.
\end{observation}
We experimented with this upper bound as an alternative to the matching score in \S\ref{ssec:results-accuracy} to show greater adaptability of our approximate-based methods. 

The streaming nature of the PS method also allows us to maintain a matching as the pair of sets are generated in the set similarity join work flow and  we describe this integration  with the verification procedure in Algorithm \ref{alg:ps_verification}. We note that the same graph building process occurs as in Algorithm \ref{alg:verify}, except the single edge generated is immediately considered for the candidate stack $Stk$ in Line \ref{alglinePSV:stream_process} of Algorithm \ref{alg:ps_verification}. This allows us to discard edges that are not pertinent to our matching as we calculate similarity scores. Following this, we finalize the matching in the post processing scheme of Line \ref{alglinePSV:post_process}, building the matching from the edges in $Stk$ and generating a score based on the weight.

\section{Evaluations}
\label{sec:evals}


As shown in Table~\ref{tab:matching_complexity}, approximate matching is theoretically faster (and consumes less memory) than exact matching based approaches. However, the impact of this improvement on the final (end-to-end) execution time is determined by the amount of time spent in matching within the compute intensive \emph{verification phase}, which performs the crucial task of calculating the similarity between two sets (see \S\ref{sec:methods-verification}). 

\begin{table}[ht!]{
\setlength{\tabcolsep}{2pt}
\renewcommand{\arraystretch}{1.1}
\scriptsize
\centering
\caption{Matching complexity per set pair of the comparing approaches. $n$ is the number of vertices of the bipartite graph generated from the pair $(R,S)$. The memory complexity of \ajps~ assumes the pairs of sets are generated one by one.}
\label{tab:matching_complexity}
\begin{tabular}{|c|c||c|c|c|}
\hline
\multirow{3}{*}{\textbf{\begin{tabular}[c]{@{}c@{}}\tokenjoin \\ (TJPJ) \cite{TokenJoinOG}\end{tabular}}} & \textbf{Variants}                                                                           & \textbf{Approx.}     & \textbf{Time}                             & \textbf{Memory}                     \\ \cline{2-5} 
       & \begin{tabular}[c]{@{}c@{}}Hungarian \\ (\tjhg)\end{tabular}                 & \multirow{2}{*}{1}   & \multirow{2}{*}{$\mathcal{O}(n^3)$}       & \multirow{2}{*}{$\mathcal{O}(n^2)$} \\ \cline{2-2}
       & \begin{tabular}[c]{@{}c@{}}Efficient \\ Verification \\ (\tjev)\end{tabular} &                      &                                           &                                     \\ \hline \hline
\multirow{3}{*}{\textbf{\begin{tabular}[c]{@{}c@{}}\apj \\ (AJ)  \\ {[}this work{]}\end{tabular}}}       & \begin{tabular}[c]{@{}c@{}}Locally \\ Dominant \\ (\ajld)\end{tabular}       & \multirow{2}{*}{0.5} & \multirow{2}{*}{$\mathcal{O}(n^2\log n)$} & \multirow{2}{*}{$\mathcal{O}(n^2)$} \\ \cline{2-2}
                                                                                                               & \begin{tabular}[c]{@{}c@{}}Greedy \\ (\ajgd)\end{tabular}                    &                      &                                           &                                     \\ \cline{2-5} 
                                                                                                               & \begin{tabular}[c]{@{}c@{}}Paz \\ Schwartzman \\ (\ajps)\end{tabular}        & 0.5                  & $\mathcal{O}(n^2)$                        & $\mathcal{O}(n \log n)$             \\ \hline
\end{tabular}
}
\end{table}

We have extended the open-source Python codebase of a state-of-the-art implementation of set similarity join (that uses exact matching), \tokenjoin~\cite{TokenJoinOG}.\footnote{\label{fn:tokenjoinfn}\href{https://github.com/alexZeakis/TokenJoin/tree/main}{https://github.com/alexZeakis/TokenJoin/tree/main}} Our code\footnote{\label{fn:sasimi}\href{https://github.com/mmandulak1/sasimi}{https://github.com/mmandulak1/sasimi}} is widely available and provided. For our experiments, we use the \tokenjoin~ methods for the candidate generation and refinement with the Positional and Joint filters (\tjpj) with a similarity threshold of $\delta=0.7$ (see threshold-based set similarity join problem statement~\ref{problem:tokenjoin}). We note that all of our experimental runs are self joins on the respective dataset.  In instances where $|\mathcal{R}|\ne 100\%$, a random sample at the given size is generated and used for all methods at that size. Our tests focus on Jaccard similarity (JAC) with an additional test using Normalized Edit similarity (NEDS) to accommodate direct string comparison. Depicted execution times are averaged over several runs with exclusive allocation on the testbed platform.

In the rest of this section, we empirically compare approximate matching (our work) with exact matching (existing state-of-the-art, \tokenjoin) based approaches for set similarity join, after briefly discussing the datasets and experimental platform. \S\ref{ssec:results-breakup} sets the expectation in terms of the achievable performance improvements across varied datasets. In \S\ref{ssec:results-baseline}, we discuss baseline performances in terms of the overall execution and verification times. We discuss the accuracy of our approximate matching based implementations in \S\ref{ssec:results-accuracy}, and establish trade-offs to improve the overall accuracy. We perform extra experiments in \S\ref{ssec:results-memory} and \S\ref{ssec:results-parallel}, showing that
our approaches consume less memory (29\% on average), alongside the parallelism potential of our approaches at a baseline of 8$\times$ improvement.

\paragraph*{\textbf{Datasets.}}
We experiment on 10 different datasets, whose characteristics are highlighted in Table \ref{tab:data}. Each token is generated by splitting words into $q$-grams ($q=3$) and substituting an integer for each $q$-gram. For the purpose of results classification, we refer to \textit{dense} instances as datasets with >100 element per set averages or >1,000 maximum element per sets. We use Jaccard (JAC) and\slash or Normalized Edit (NEDS) similarity measures (see \S\ref{background}).
\begin{table}[!ht]
{\footnotesize
\scriptsize
\centering
\caption{Experimental datasets and their characteristics.}
\label{tab:data}
\begin{tabular}{|l|c||c|cc|}
\hline
\multirow{2}{*}{\textbf{Datasets}} & \multirow{2}{*}{\textbf{Elements $\in$ Set}} & \multirow{2}{*}{\textbf{\#Sets}} & \multicolumn{2}{c|}{\textbf{Ele/Set}} \\ \cline{4-5}
                                   &                                              &                                  & \multicolumn{1}{c|}{\textbf{Avg}} & \textbf{Max} \\ \hline \hline
LIVEJ~\cite{LiveJ-Orkut}\footnote{\label{fn:socialnet}\href{http://socialnetworks.mpi-sws.org/data-imc2007.html}{http://socialnetworks.mpi-sws.org/data-imc2007.html}} & interests $\in$ user                         & 3.1M & \multicolumn{1}{c|}{36} & 300  \\ \hline
AOL~\cite{AOL}\linkfoot{https://www.cim.mcgill.ca/~dudek/206/Logs/AOL-user-ct-collection/} & keywords $\in$ search                        & 1.8M & \multicolumn{1}{c|}{3}  & 73   \\ \hline
KOSARAK~\cite{kosarak}\linkfoot{http://fimi.uantwerpen.be/data/} & links $\in$ user                     & 610K & \multicolumn{1}{c|}{12} & 2.5K \\ \hline
ENRON~\cite{TokenJoinOG}\footnote{\label{fn:tokenJoinData}\href{https://github.com/alexZeakis/TokenJoin/tree/main}{https://github.com/alexZeakis/TokenJoin/tree/main}} & words $\in$ email                            & 518K & \multicolumn{1}{c|}{134} & 3.2K \\ \hline
DBLP~\cite{TokenJoinOG} 
\cref{fn:tokenJoinData} & authors $\in$ work & 500K & \multicolumn{1}{c|}{13} & 189 \\ \hline
FLICKR~\cite{TokenJoinOG}
\cref{fn:tokenJoinData} & words $\in$ photo & 500K & \multicolumn{1}{c|}{9} & 361 \\ \hline
GDELT~\cite{TokenJoinOG}
\cref{fn:tokenJoinData} & topics $\in$ article & 500K & \multicolumn{1}{c|}{19} & 396 \\ \hline
BMS-POS~\cite{BMS-POS}\linkfoot{https://www.kdd.org/kdd-cup/view/kdd-cup-2000} & items $\in$ sale & 320K & \multicolumn{1}{c|}{6} & 164 \\ \hline
YELP~\cite{TokenJoinOG}
\cref{fn:tokenJoinData} & types $\in$ business & 160K & \multicolumn{1}{c|}{6} & 47 \\ \hline
MIND~\cite{TokenJoinOG}
\cref{fn:tokenJoinData} & words $\in$ article & 123K & \multicolumn{1}{c|}{32} & 357 \\ \hline
\end{tabular}
}
\end{table}

\paragraph*{\textbf{Platforms.}}
Our baseline results are collected using a single thread on a system with 2TB of DDR4 RAM and dual-socket NUMA AMD EPYC 7742 2.2GHz 64-core processors and 2 threads\slash core with Ubuntu version 20.04.4 LTS O/S. We use Python version 3.12.9 and \tjhg uses NetworkX version 3.4.2. For the preliminary parallel experiments (Appendix ~\ref{ssec:results-parallel}), GNU C++ compiler (v13.3) with OpenMP 5.1 were used.



\subsection{Timing Breakdown}\label{ssec:results-breakup} 
\label{sec:runtimeDist}
\begin{figure}[!ht]
    \centering
    \includegraphics[scale=0.52]{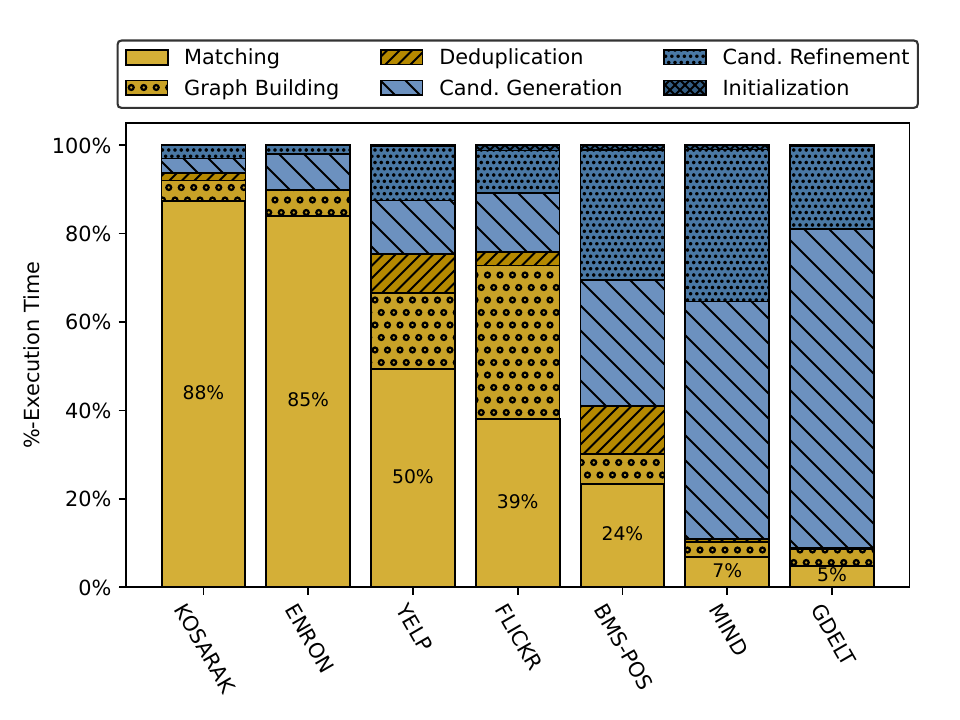}
    \caption{Execution time  distribution for runs of \tjhg at $|\mathcal{R}|=\text{50K}$. The annotated percentage is the exact value for matching, rounded up. Verification includes Matching, Deduplication and Graph Building steps.}
    \label{fig:verifyTimeDist}
\end{figure}
Using 50K samples\slash dataset, we show in Fig.~\ref{fig:verifyTimeDist}, the percentage breakdown of the different components from the \tjpj workflow, comprising of 
initialization, candidate pairing generation, refinement and verification (deduplication, graph building and subsequent matching combined). For Fig.~\ref{fig:verifyTimeDist}, we consider the datasets for which the verification phase requires more than 5\% of the overall execution time. 
We observe high variations in the verification related computations (between <1--94\% of the overall execution times). Within the verification phase, matching is typically the most computationally expensive component. Generally larger datasets with higher number of elements\slash set tend to spend more time on matching, with the exception of the MIND and GDELT datasets. This disparity can be explained by examining the average number of elements\slash set of the candidate pairings that proceed through the verification phase. For instance, GDELT with $|\mathcal{R}|=\text{5K}$ produces up to 2M candidate pairings, out of which only 11K proceed through the verification following refinement. Within the 11K pairings, average elements\slash set is about 4; the resultant graph will have low-degree edges, significantly reducing the matching overheads.

\begin{figure}[!ht]
\hspace*{-0.48cm} 
    \centering
    \includegraphics[scale=0.32]{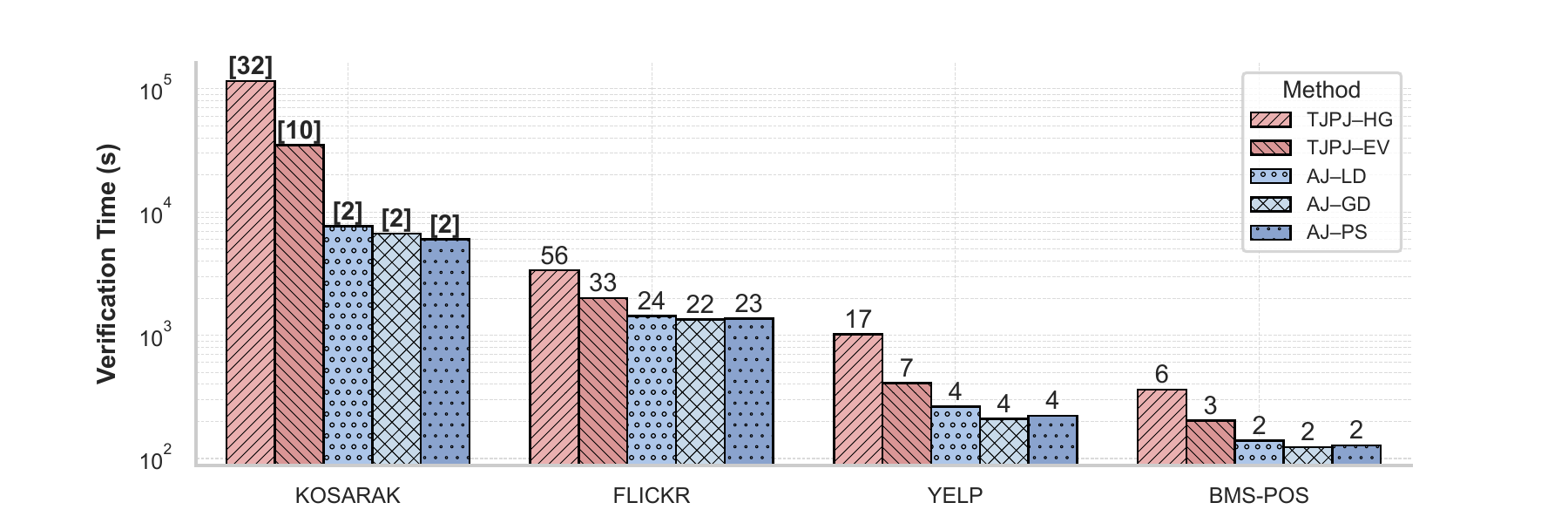}
    \caption{JAC verification execution time comparison per matching method, $|\mathcal{R}|=\text{100\%}$. The [bold] annotations depict time in hours, while the rest depict time in minutes, rounded up to the nearest whole number.}
    \label{fig:jaccardRes}
\end{figure}

\subsection{Baseline Performance}\label{ssec:results-baseline}
In this section, we discuss performance results of the various approximate matching methods compared to exact Hungarian and Efficient-Verification in original \tokenjoin~ (see Table~\ref{tab:matching_complexity}), in terms of the total execution and verification times using Jaccard (JAC) and NEDS similarity measures (in \S\ref{sssec:jac_verification_runtime}, \S\ref{sssec:jac_total_runtime} and \S\ref{sssec:neds_verification_runtime}). We also discuss scalability (\S\ref{sssec:jac_scalability}), effect of threshold variation (\S\ref{sssec:jac_threshold}) and finally summarize the performance aspects in \S\ref{sssec:perf_summary}.

\subsubsection{\textbf{JAC verification time:}}\label{sssec:jac_verification_runtime} 
Fig.~\ref{fig:jaccardRes} exhibits the execution time of the verification phase using Jaccard  similarity within \tjpj ($\delta=0.7$) at $|\mathcal{R}|=\text{100\%}$.  In all instances, the approximate methods yield faster verification than the original \tjhg, with the most improvements in KOSARAK and YELP, with 19.1$\times$ improvement using \ajps and 4.8$\times$ using \ajgd, respectively. Across all datasets, we see a geometric mean improvement over \tjhg of 4.5$\times$ for \ajps and \ajgd, with 3.9$\times$ for \ajld. The arithmetic mean for each is 6$\times$ for \ajps, 5.7$\times$ for \ajgd and 5$\times$ for \ajld. Compared to \tjev, we continue to observe faster execution times for the approximation methods. The biggest improvement is seen using \ajps on KOSARAK, with 5.8$\times$ performance improvement relative to \tjev, with the next best being DBLP with 2.3$\times$ improvement with \ajps. On an average, \ajgd depicts 2.4$\times$ improvement, \ajld demonstrates 2.1$\times$ improvement and \ajps shows 2.5$\times$ improvement, relative to \tjev. In terms of the verification times, \ajgd performs the best in 3\slash 6 cases, while \ajps performs equally better for the rest. Individual differences between the approximation methods is marginal in most cases, except for KOSARAK (where we observed the best improvement using \ajps, about 19$\times$ relative to \tjhg), where the most time difference between \ajgd and \ajps is about a quarter of an hour. Otherwise, the geometric mean percentage difference of maximum and minimum verification times between proposed approximation methods is about 16\%. Thus, we see a wide range (1.9--19.1$\times$) of performance improvements in the verification process when comparing the approximation methods to \tjhg and to \tjev. 
\begin{figure}[!ht]
\hspace*{-0.75cm} 
    \centering
    \includegraphics[scale=0.34]{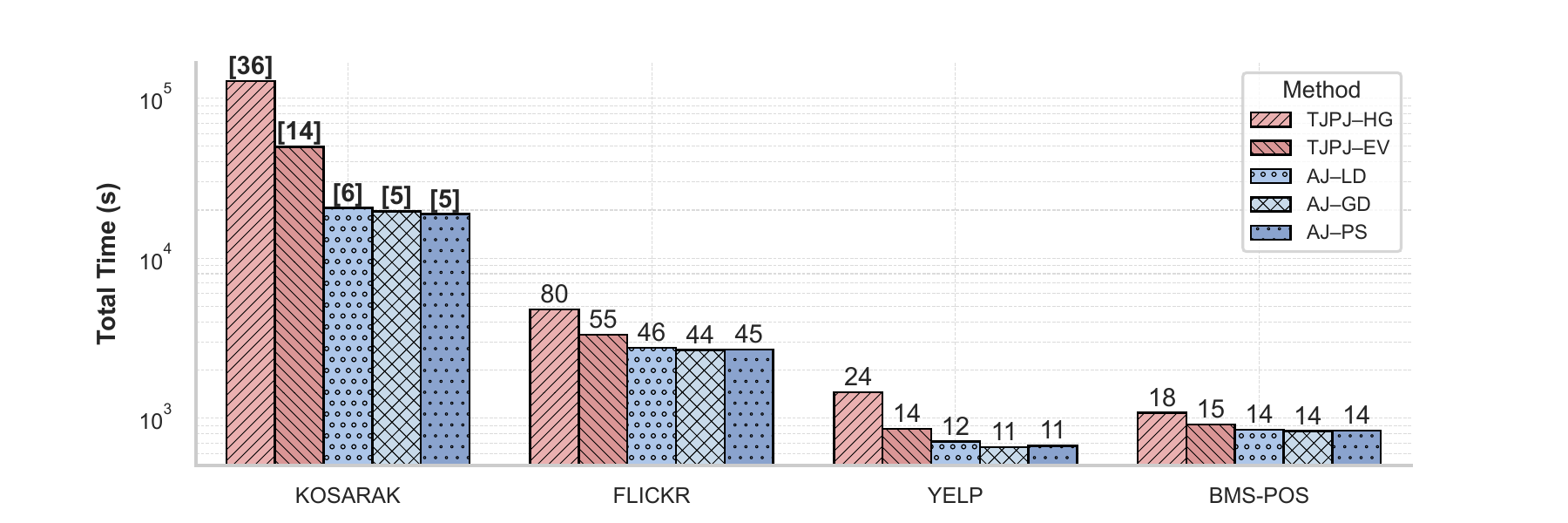}
    \caption{JAC total execution time comparison per matching method, $|\mathcal{R}|=\text{100\%}$. The [bold] annotations depict time in hours, while the rest depict time in minutes, rounded up to the nearest whole number.}
    \label{fig:jaccardTotal}
\end{figure}
\subsubsection{\textbf{JAC total execution time:}}\label{sssec:jac_total_runtime}
We show the total execution time of the fuzzy set similarity join workflow in Fig.~\ref{fig:jaccardTotal}. In every case, approximation methods outperform \tjhg and \tjev, with performance improvements being 3.78$\times$ compared to \tjhg and 2.18$\times$ vs. \tjev on average. Not surprisingly, cases where the verification time is relatively high, we observe substantial impact on the total execution time (refer to \S\ref{ssec:results-breakup}). Specifically, KOSARAK (94\% of the total time spent in verification) sees the highest improvements relative to \tjhg: 6.5$\times$ with \ajgd, 6.2$\times$ with \ajld and 6.7$\times$ with \ajps, respectively. In terms of actual execution times, this equates to a difference of more than a day's worth of computing (about 30 hours)! Depending on the dataset density and the verification times, the improvements are more impactful for the larger instances. YELP exhibits the second highest performance, depicting 2.1$\times$ improvement in the execution times across the approximate methods. In general, we observe about 2.2$\times$ improvement in the (total) execution times across the datasets against \tjhg. On the other hand, compared to \tjev, there is a 2.5$\times$ improvement for KOSARAK and an average 1.4$\times$ improvement for the approximation methods considering rest of the datasets. Thus, we see a range of 1.4--6.7$\times$ performance improvement with the approximate methods (\ajld, \ajgd and \ajps) as compared to exact \tjhg and \tjev. 

\subsubsection{\textbf{JAC execution time scaling:}}\label{sssec:jac_scalability}
\begin{figure}[!ht]
    \centering
    \includegraphics[scale=0.45]{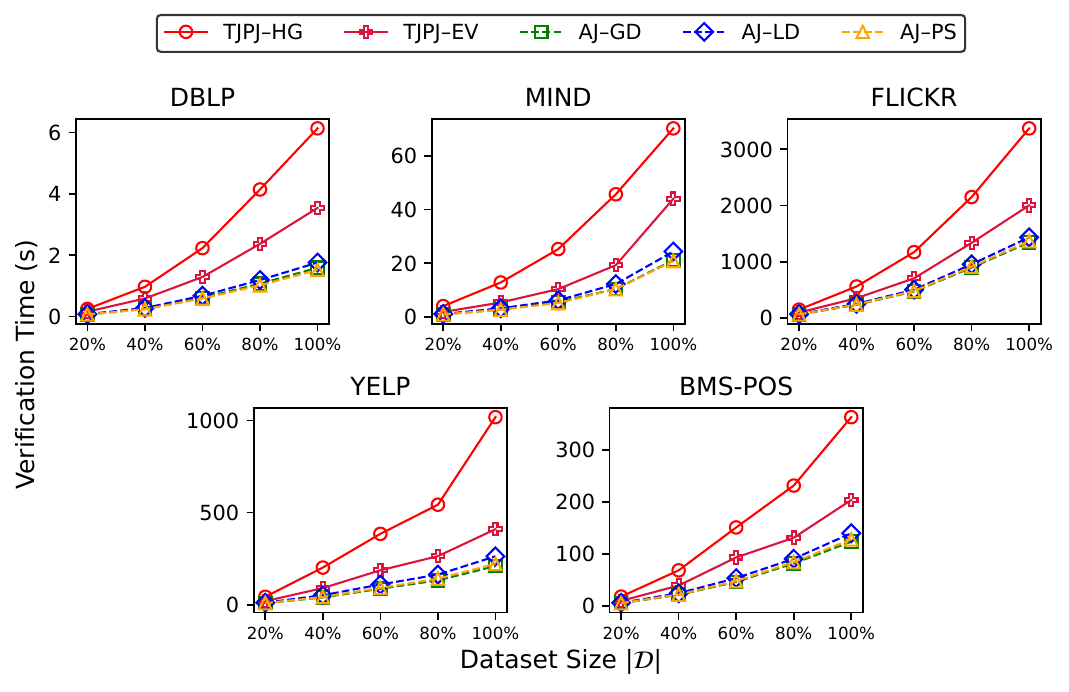}
    \caption{JAC verification execution time scaling per set size (lower is better). Average improvement of 3.4$\times$ vs. \tjhg and 1.8$\times$ vs. \tjev.}
    \label{fig:jacRuntimeScale}
\end{figure}

In Fig.~\ref{fig:jacRuntimeScale}, we perform scaling experiments on a subset of our datasets to assess the performance as the data sizes increase, taking random samples of each dataset at 20\% size intervals. 
\begin{figure}[!ht]
    \centering
    \includegraphics[scale=0.40]{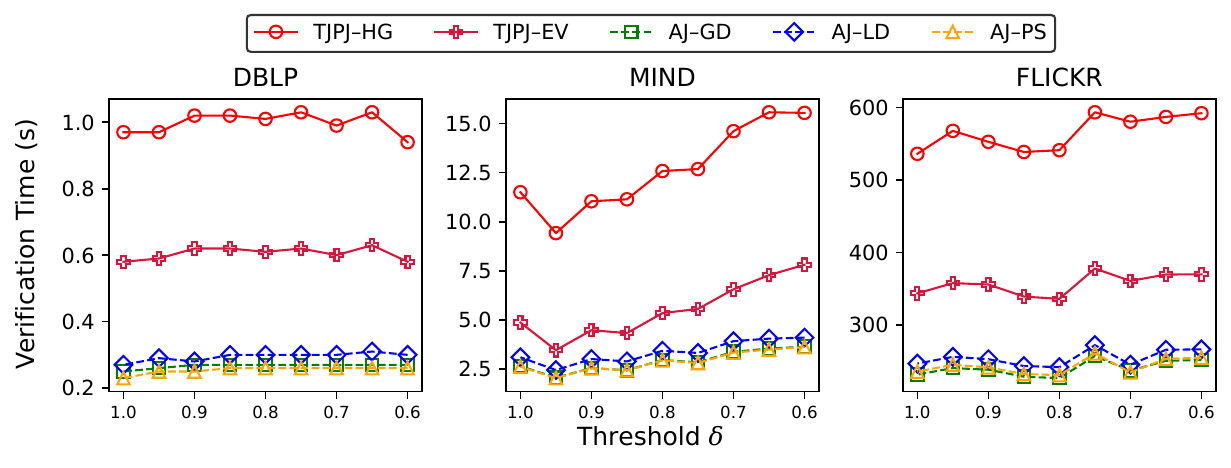}
    \caption{JAC threshold variation verification execution time scaling (lower is better). Average performance improvement of 3.4$\times$ vs. \tjhg and 1.8$\times$ vs. \tjev.}
    \label{fig:jacThresholdVar}
\end{figure}
Across every instance, we observe improved scaling trends with reduced execution times for the approximation methods at each size interval. Best performance is achieved at 80\%--100\% for FLICKR, 4.9$\times$ better performance using \ajgd than \tjhg, and, 2.5$\times$ using \ajps as compared to \tjev. On average, the approximation methods shows 3.4$\times$ improvement relative to \tjhg and 1.8$\times$ improvement against \tjev. We see \ajgd outperforming the rest in 16\slash 25 instances, with \ajps exhibiting better performance for the remaining 9\slash 25, with an average \%-difference of 14\% in terms of execution times among the approximation methods. 

\subsubsection{\textbf{NEDS verification time:}}\label{sssec:neds_verification_runtime} 
\begin{figure}[!ht]
\hspace*{-0.70cm} 
    \centering
    \includegraphics[scale=0.32]{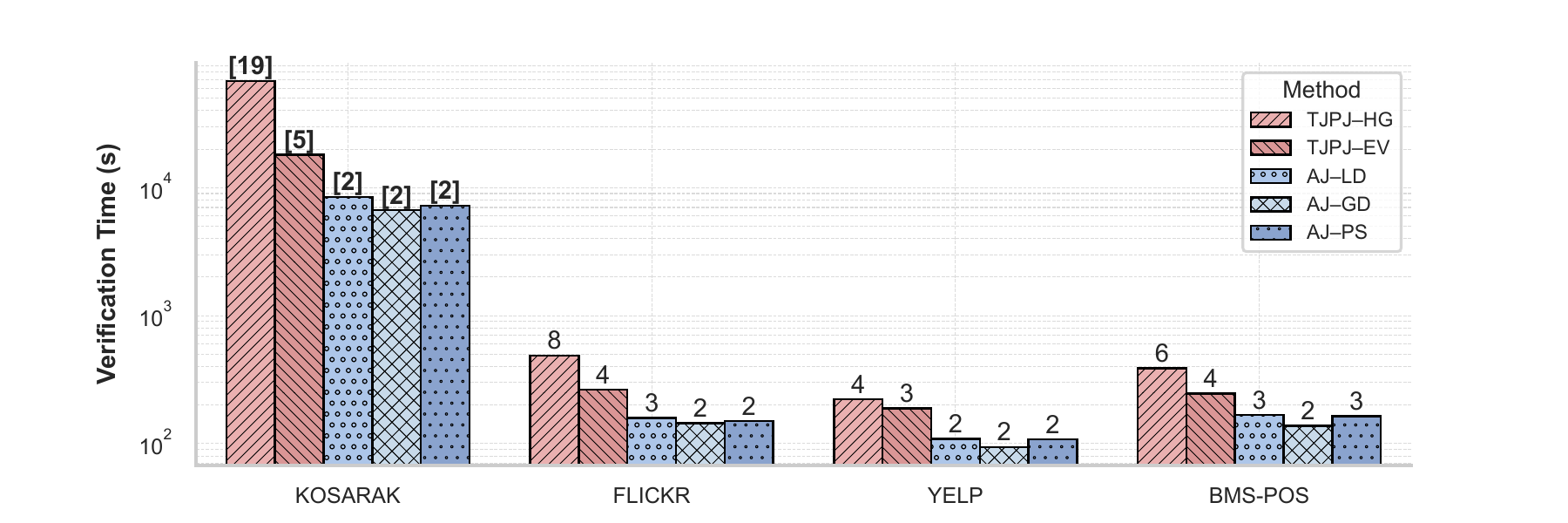}
    \caption{Verification time using NEDS at $|\mathcal{R}|=\text{20\%}$. 3.7$\times$ improvement on average against \tjhg and 1.8$\times$ against \tjev. The [bold] annotations depict time in hours, while the rest depict time in minutes, rounded up to the nearest whole number.}
    \label{fig:neditRes}
\end{figure}

We also test the performance using NEDS on a subset of our datasets due to significantly higher execution times as compared to JAC. For this scenario, we report the verification times on $|\mathcal{R}|=\text{20\%}$ using \tjpj ($\delta=0.7$) as our default. The results are reported in Fig.~\ref{fig:neditRes}. We observe approximate matching cases outperforming the default across every instance, with KOSARAK seeing the most improvement of 10.2$\times$ using \ajgd compared with \tjhg. Consequently, \ajgd depicts the highest performance improvement of 2.7$\times$ against \tjev. Across every instance, we observe an arithmetic mean improvement of 3.7$\times$ against \tjhg and 1.8$\times$ as compared to \tjev. In general, \ajgd performs the best across all the 5 instances of the approximation methods (as shown in Fig.~\ref{fig:neditRes}), depicting a performance variation of up to 14\% among the approximation methods considered.

\subsubsection{\textbf{JAC threshold variation:}}\label{sssec:jac_threshold} 
We also experiment with varying $\delta$ to ensure execution time improvements are consistent relative to candidate set sizes. We present only the significant results in Fig.~ \ref{fig:jacThresholdVar}. Throughout the datasets, we observe consistent execution time trends with constant performance improvement gaps between the \tjhg and \tjev methods, and the approximate methods. The highest performance improvement on average is observed at the lower values of $\delta$, as there are more candidate pairings for verification improvement. Considering the approximate methods, we observe on an average 3.4$\times$ improvement upon \tjhg across all $\delta$ values and 1.8$\times$ improvement upon \tjev. Among the 45 instances considered, \ajgd performed the best in 27 while \ajps performed the best in the other 18. 


\begin{figure}
    \centering
    \includegraphics[scale=0.65]{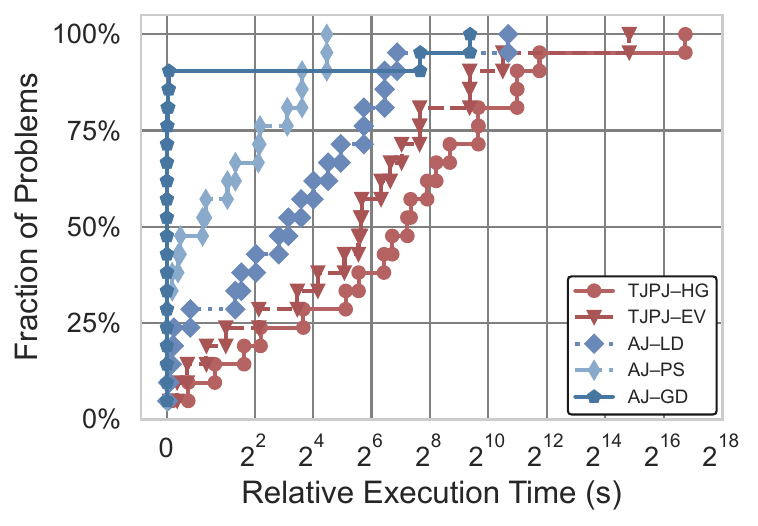}
    \caption{A performance profile to summarize the execution time differences between \apj~methods compared to \tokenjoin. The X-axis (log scale) represents the  execution time in seconds, normalized by the best performing algorithm for a given problem. We normalize the time by subtracting the time taken by the best performing method from all the methods. The Y-axis represents the fraction of the problems. The closer to the left, the better performance.}
    \label{fig:perfPlotResults}
\end{figure}

\subsubsection{\textbf{Performance summary:}} \label{sssec:perf_summary}
\begin{table}[!ht]
\centering
\caption{Summary of relative performance improvements (approx. vs. exact) as geometric mean: higher is better.}
\label{tab:sum_stats}
\begin{tabular}{@{}lcccc@{}}
\toprule
\multirow{2}{*}{Method} &
  \multicolumn{2}{c}{\textbf{Verification Time}} &
  \multicolumn{2}{c}{\textbf{Total Time}} \\
  \cmidrule(lr){2-3} \cmidrule(lr){4-5}
  & \textbf{HG} & \textbf{EV} & \textbf{HG} & \textbf{EV} \\
\midrule
\ajld & 4.27$\times$ & 2.00$\times$ & 2.43$\times$ & 1.46$\times$ \\
\ajgd & 4.70$\times$ & 2.20$\times$ & 2.49$\times$ & 1.51$\times$ \\
\ajps & 4.88$\times$ & 2.29$\times$ & 2.56$\times$ & 1.55$\times$ \\
\bottomrule
\end{tabular}
\end{table}

The overall performance improvements observed by the approximation methods relative to the exact state-of-the-art are summarized in Table~\ref{tab:sum_stats} and Fig.~\ref{fig:perfPlotResults}. Table~\ref{tab:sum_stats} shows the geometric mean of performance improvements of approximate methods relative to exact counterparts.
We observe an average of 2$\times$ improvement in favor of approximate methods against both \tjhg and \tjev for the total execution times, with 2--4$\times$ improvement for the verification times. The relative differences between verification and total time improvements relies on the overall distribution of the verification times, as discussed in \S\ref{sec:runtimeDist} and \S\ref{sssec:jac_total_runtime}. 

Fig.~\ref{fig:perfPlotResults} shows the compiled relative performance of all of our test instances, relative to the best method as a difference of execution time. For each test instance, the best algorithm (i.e., the algorithm with the least execution time) is set to zero, and all others' execution time is offset from the best one.  Approximate methods outperform the exact methods in every instance. In general, we see the strongest performance results being competitive between \ajps and \ajgd, with a slight performance advantage on average for \ajps despite \ajgd having the best execution times in a majority of cases. We expand on the accuracy benefits of the approaches in the forthcoming section, \S\ref{ssec:results-accuracy}.

\subsection{Approximate Matching Accuracy}\label{ssec:results-accuracy}

\begin{table*}[!ht]
\scriptsize
\centering
\caption{Accuracy assessment of our approximate matching based approach, $|\mathcal{D}|=\text{50K}$; Recall\slash Precision closer or equal to 1 is ideal.}
\label{tab:recall_precision}
\renewcommand{\arraystretch}{1.2} 
\setlength{\tabcolsep}{6pt} 
\begin{tabular}{l|ccc|ccc||ccc|ccc}
\hline
\multirow{2}{*}{\textbf{Dataset}} & 
\multicolumn{6}{c||}{\textbf{Matching Weight based (\apj)}} & 
\multicolumn{6}{c}{\textbf{Upper bounds based (\apj)}} \\ \cline{2-13} 
& \multicolumn{3}{c|}{\textbf{$\Delta$\#Sets}} & 
  \multicolumn{3}{c||}{\textbf{Recall}} & 
  \multicolumn{3}{c|}{\textbf{$\Delta$\#Sets}} & 
  \multicolumn{3}{c}{\textbf{Precision}} \\ \cline{2-13}
& \textbf{GD} & \textbf{LD} & \textbf{PS} & 
  \textbf{GD} & \textbf{LD} & \textbf{PS} & 
  \textbf{GD} & \textbf{LD} & \textbf{PS} & 
  \textbf{GD} & \textbf{LD} & \textbf{PS} \\ \hline \hline
LIVEJ   & 0     & 0     & 0     & 1.0000 & 1.0000 & 1.0000 & +2    & +2    & +1    & 0.9998 & 0.9998 & 0.9999 \\
AOL     & 0     & 0     & 0     & 1.0000 & 1.0000 & 1.0000 & 0     & 0     & 0     & 1.0000 & 1.0000 & 1.0000 \\

KOSARAK & -1    & -7    & -1    & 0.9999 & 0.9999 & 0.9999 & +4    & +4    & +4    & 0.9999     & 0.9999     & 0.9999    \\
ENRON   & -4515 & -4637 & -593  & 0.9293 & 0.9274 & 0.9757 & +835    & +835    & +778    & 0.9871     & 0.9871     & 0.9879     \\
DBLP    & 0     & 0     & 0     & 1.0000 & 1.0000 & 1.0000 & +1    & +1    & +1    & 0.9870     & 0.9870     & 0.9870    \\
FLICKR  & -115  & -115  & -91   & 0.9994 & 0.9994 & 0.9995 & +114    & +114    & +109    & 0.9994     & 0.9994     & 0.9995      \\
GDELT   & -937  & -937  & -481  & 0.9992 & 0.9992 & 0.9996 & +843    & +843    & +806    & 0.9992     & 0.9992     & 0.9993     \\
BMS-POS & 0     & 0     & 0     & 1.0000 & 1.0000 & 1.0000 & 0    & 0    & 0    & 1.0000     & 1.0000     & 1.0000     \\
YELP    & -47   & -47   & -80   & 0.9999 & 0.9999 & 0.9998 & +293    & +293    & +199    & 0.9999     & 0.9999     & 0.9999     \\
MIND    & -38   & -38   & -8    & 0.9995 & 0.9995 & 0.9999 & +14    & +14   & +13    & 0.9998    & 0.9998     & 0.9998     \\ \hline
\end{tabular}
\end{table*}

Minor variations in the matching weights are expected for the approximate methods; we consider the exact Hungarian method as our ground truth in assessing the accuracy of approximate matching based verification.
Since both \tjhg and \tjev produces the same matching, their qualities are exactly similar.  We calculate the resultant set size of the approximate matching verification and compare that to the default, checking for discrepancies in the set sizes ($\Delta\#\text{Sets}$) to calculate recall and precision metrics. We define recall and precision relative to our ground truth based on standard true positives (TP), false positives (FP) and false negatives (FN) formulations:
\begin{equation*}
    Recall=\frac{TP}{TP+FN},~~Precision=\frac{TP}{TP+FP}.
\end{equation*}
 We denote $TP$ as instances in the resultant set for both the Hungarian and approximate matching verifications, $FP$ as the extra instances not found in the original (resultant) set, and $FN$ as the missing instances.
 
 \subsubsection{\textbf{Matching weight based accuracy:}} We present the default results in the first half of Table~\ref{tab:recall_precision}, using the \emph{lowest} values from 3 runs of random samples of $|\mathcal{R}| =\text{50K}$. From these runs, we conclude that the usage of approximate matching yields recall values >0.9 across all instances, with all datasets having at least 0.99 recall besides ENRON. We note that we have excluded the precision as all datasets yielded a precision of 1.0. This implies that our datasets are subsets of the dataset yielded by the \tjhg method with an average of approximately 570 pairings missing for \ajgd and \ajld, and 120 pairings for \ajps. \ajps also shows the highest recalls among all methods, having the highest value when not tied by the other approximation methods for all 10 datasets. By geometric mean, we see an average recall of \textbf{0.9824, 0.9822} and {\bf 0.9974} for \ajgd, \ajld and \ajps, respectively. We attribute the exception of ENRON statistics to its high matching usage (about 87\%) alongside a large amount of candidate sets (about 70K) and relatively high dataset density (average 133 elements\slash set). Nevertheless, we can obtain significant performance improvements of up to {\bf 6$\times$} while maintaining a majority of the resultant pairings with {\bf 99\%} recall. 

\subsubsection{\textbf{Upper bound based accuracy:}} To improve the recall metric using approximate matching, we experiment with trade-offs regarding accuracy by using upper bounds generated from the approximate matching for verification. Since \ajgd and \ajld are half-approximate solutions, we simply double the matching weights, whereas for the \ajps method, we use the sum of $\phi$ from the primal-dual basis discussed in section~\ref{sec:ps} (Observation~\ref{obs:dual-bound}). We report the lowest values from 3 runs of random samples considering $|\mathcal{R}| =\text{50K}$ in the second half of Table~\ref{tab:recall_precision}. We omit the associated recall values, as every dataset instance led to a 100\% recall, implying that the resultant sets include the default\slash exact method's pairings, in addition to new pairings. This inclusion of the extra pairings is represented by the precision metric for each method (in Table~\ref{tab:recall_precision}); geometric mean of precision across the approximate matching methods are \textbf{0.9978, 0.9978} and {\bf 0.9980} for \ajgd, \ajld and \ajps, respectively. We observe that shifting the bounds leads up to a thousand extra pairings in the resultant dataset while including the ones missed in the first portion of Table \ref{tab:recall_precision}. On an average, \ajgd and \ajld produces 210 extra pairings while \ajps yields 191 across the datasets considered. Nevertheless, we observe high precision values (about 0.98) using proposed approximate matching methods within verification.
Based on these accuracy assessments, we can consider adapting the approximate matching methods to yield more or less pairings in the resultant dataset. Specifically, if retaining the pairings generated by an exact method such as \tjpj is essential, then shifting the bounds can induce the necessary pairings without affecting the overall performance (experiments indicate less than 2\% performance loss upon expanding the approximation bounds). However, this type of adaptation would not be necessary if the application can withstand minor loss of pairings, i.e., taking a mild hit at the accuracy. Regardless of the choice, we demonstrate significant performance improvements of {\bf 1.2--6.1$\times$} with high and acceptable accuracy values.




\section{Related Work.}
\label{sec:related}
Set similarity and fuzzy set similarity problems have been studied extensively.
We briefly discuss the overarching processes and their related works here. In \S\ref{ssec:opt-match}-\ref{ssec:stream-match}, we provide relevant discussion on optimal, approximate, and semi-streaming matching algorithms.

\subsection{\textbf{Traditional Set Similarity:}} Traditional and exact set similarity works have been studied as a fundamental problem in data processing \cite{setsimApp1,setsimApp2,setsimApp3,setsimApp4,setsimApp5,setsimSurvey1}, often utilizing a method based on candidate pair generation and verification through a filter-verify framework. These methods utilize similarity functions (e.g., Jaccard, cosine, or overlap coefficient\cite{simMeasuresSurvey1}) to determine set overlap. Typical advancements rely on improving filtering techniques that consider tradeoffs between efficiency and effectiveness \cite{filteringMethod1,filteringMethod2} and targeting performance improvements in large-scale applications. Filtering methods commonly studied include \textit{prefix filtering} \cite{prefixFilter}, \textit{size filtering} \cite{sizeFilter} and \textit{positional filtering} \cite{positionalFilter}, which have all seen respective modern improvements. Traditional set similarity methods are not robust to perturbations within data, often leading to imprecise similarity values when data is slightly ambiguous. These cases are better considered in the \textit{fuzzy set similarity} to approximately match set elements.

\subsection{\textbf{Fuzzy Set Similarity:}} 
The fuzzy set similarity join problems are based on bipartite-matching that approximates the set relatedness. Matching algorithms are computationally expensive, which increases the verification time, thus encouraging advanced filtering methods to outweigh the slower runtimes in the filter-verify framework. We discuss some of the previous methods that extend this motivation, while focusing on optimizations within verification.

The early work includes the FastJoin method \cite{FastJoin}, which proposes a fuzzy-token similarity function to address the fuzzy set similarity problem. This method works by assigning {\em signatures} to token sets generated as $q$-grams, identifying sets containing those signature tokens and performing the bipartite matching to compute fuzzy overlap. They also introduce the concept of a threshold for the matching weight relative to set similarity and apply the filter-verify framework in the context of fuzzy set similarity. 

SilkMoth~\cite{SilkMoth} checks for approximate equivalence of sets through set similarity and further checks for subset properties as set containment. Like FastJoin, SilkMoth follows a signature-based scheme but eliminates false-negatives through filtering to better curate their candidate set. They further improve verification by considering the triangle inequality of mapping the similarity function to a distance function $\psi(r,s)=1-\phi(r,s)$. Thus, if the triangle inequality is satisfied by $\psi$, then the identical elements of $R$ and $S$ must be in the maximum matching. This reduces the number of vertices in the bipartite graph, as they can be immediately inserted into the matching beforehand. SilkMoth showed 13$\times$ performance relative to FastJoin with a 30-50\% improvement in verification times using the optimization.

The \tokenjoin~ method \cite{TokenJoinOG} improves SilkMoth and the state-of-the-art for fuzzy similarity joins. \tokenjoin~ relies on a token-based filtering method (rather than the element-based filtering of SilkMoth), which assigns utility values to tokens and optimizes filtering relative to these utilities. For this, they include a \textit{positional filter} and a \textit{joint utility filter}, utilizing token positions in prefixing and utilities between the set pairing $(R,S)$ for improved refinement, respectively. The \tokenjoin~ method also introduces an \textit{Efficient Verification} technique that can terminate the Hungarian algorithm early based on current upper and lower bound of the matching and the given threshold.
Experiments show up to a 4.5$\times$ performance improvement of Efficient Verification on large dataset instances with element counts greater than 100. Results demonstrate that \tokenjoin~ is an order of magnitude faster than SilkMoth.

\section{Concluding Remarks}
In this paper, we devise \apj, a fuzzy set similarity join method using approximate maximum weighted matching. We implement three approximate matching methods within candidate verification: the Greedy (\ajgd), Locally-Dominant (\ajld) and Paz Shwartzman (\ajps) methods, and comprehensively compare the execution time performance of our method to the current state-of-the-art, \tokenjoin. Across a variety of sparse and dense datasets, \apj~ outperforms both the optimal (\tjhg) and efficient verification (\tjev) matching methods of \tokenjoin~ by 2-19$\times$ with high accuracy (0.99 recall on average). We also show that \ajps uses 23\% less memory on average than \tjhg and \tjev (Appendix \S\ref{ssec:results-memory}).

Extending our work to parallel implementations of approximate matching-based verification, our proxy instance shows the potential for high performance gains (a baseline of 8$\times$ using base parallel methods) with nuanced parallelization schemes, as discussed in Appendix \S\ref{ssec:results-parallel}. Potential avenues to improve performance in this aspect exist through extra memory allocations (distributed stacks), improved workload distributions (data size-dependent), or even advanced parallel platforms (GPUs). In each case, there is a notable tradeoff in complexity and resource usage to see significant performance gains, thus generalizing the challenges faced as those explored in works of parallel prototypical graph problems. We note such a tradeoff analysis in parallel verification schemes as an avenue of extension.



\clearpage
\bibliographystyle{siamplain}

\clearpage
\appendix

\section{Bipartite Weighted Matching}
\label{sec:Matchings}
\subsection{LP formulation and Duality}
\label{ssec:LP}

Linear programming (LP) and duality theory play a significant role in designing algorithms for bipartite matching problems. In Eqn.~\ref{form:primal-match} of Fig.~\ref{fig:match-formulations}, we show the primal linear programming relaxation of the MWBM problem. We define $x(e)$ a non-negative real variables for each edge, which encodes the matching in the graph. The objective function is to maximize the sum of weights of the edges in the matching. This linear program relaxes binary $x$ variables in the integer linear program of MWBM, and thus provides an upper bound on the optimum but does not necessarily guarantee an integer (i.e., $x(e) \in \{0,1\})$ solution. However, for bipartite matching, the primal constraints can be expressed as $Ax \leq b$, where $A$ is an $n\times m$ incidence matrix of $G$ and $b$ is a vector of all 1s. Since $A$ is totally unimodular and $b$ is integral, there exists an integral optimal solution of the primal problem that represents a feasible maximum weighted matching.


\setlength{\columnsep}{1pt}
\begin{figure}[h]
\small
\scriptsize
\begin{multicols}{2}
\begin{maxi}
    {}{ \sum_{e \in E} w(e) x(e) }{ \label{form:primal-match}}{} 
    \addConstraint{\sum_{e \in \delta(v)} x(e)}{\leq 1}{\quad \forall v \in V}
    \addConstraint{ x(e)}{\geq 0}{\quad \forall e \in E}
\end{maxi}

\columnbreak

\begin{mini}
    {}{\sum_{v \in V} y(v)}{\label{form:dual-match}}{} 
    \addConstraint{y(u) + y(v)}{\geq w(e)}{\quad \forall \{u,v\} \in E}
    \addConstraint{y(v)}{\geq 0}{\quad \forall v \in V}
\end{mini}
\end{multicols}
\caption{Primal (left) and dual (right) formulations for the LP relaxation of the matching problem.}
\label{fig:match-formulations}
\end{figure}

For each primal problem, we can also devise a dual LP that provides an upper bound on the primal. Here in Eqn.~\ref{form:dual-match}, the dual variables $y$ are defined on each vertex and the dual constraints are derived from the standard dual formulation technique.  
We say a solution pair of both primal and dual problem a feasible pair if both of these solutions (i.e., $x$ and $y$) satisfy the constraints. We state the following result from standard duality theory of linear programming, which is used in the Hungarian algorithm and the semi-streaming PS algorithm described in section~\ref{sec:methods-matching}.

\begin{lemma}
\label{lem:duality}
    Let $(x,y)$ be a feasible primal-dual pair  and $W_*$ be the weight of a maximum weighted matching, then $\sum_{v \in V} y(v)$ $\geq$ $\sum_{e \in E} w(e) \cdot x(e) \geq W_*$ (Weak Duality). Furthermore, if $\sum_{v \in V} y(v) = \sum_{e \in E} w(e) \cdot x(e)$, and $x$ is integral (which exists since A is totally unimodular), then $x$ encodes the maximum weighted matching of $G$ (Strong Duality).
\end{lemma}


\subsection{\textbf{Optimal Matching:}}
\label{ssec:opt-match}
Matching is arguably one of the most studied combinatorial optimization problems (defined in \S\ref{ssec:bipartiteMatching}) \cite{schrijver-book}.
The Hungarian algorithm, as we know from the work of Kuhn~\cite{kuhn1955hungarian} for bipartite weighted matching, was in fact known as early as Jacobi~\cite{jocobi1865investigando}. The famous Blossom algorithm of Edmonds~\cite{edmonds1965paths} that works for general graphs  pioneered the notion of polynomial time algorithms as an efficient algorithmic paradigm. The exact algorithms, although polynomial, is both complex and expensive from a real-world dataset perspective. These motivate the development of approximate weighted matching, where instead of computing optimal matching, one seeks for a matching with weights at least $0<\alpha < 1$ times the optimal. 

\subsection{\textbf{Approximate Matching:}} 
\label{ssec:approx-match}
Developing approximate matching algorithms has been the forefront of matching research for the last few decades. Among these, the simplest and the most popular one is the 1/2-approximate greedy algorithm~\cite{avis1983survey} (Henceforth, GD), which sorts the edges according to weights in descending order and then scans the edges to compute a maximal matching in this order. The execution time and memory complexity of the greedy algorithm is $O(m \log n)$ and $O(m)$, respectively, for a graph with $n$ vertices and $m$ edges. The first linear time (i.e., $O(m)$) algorithm for 1/2-approximate matching is due to Preis \cite{PreisMatching}, which uses local dominance properties of the greedy matching. Later it was observed that these local properties are also useful in parallel computing environments. The locally dominant algorithm (Henceforth, LD), which matches a vertex to its locally heaviest available neighbor, along with its variant, are the most performant matching algorithm in shared memory~\cite{Halappanavar2012, manne2014new}, distributed memory \cite{Catalyurek2011}, and GPU architectures~\cite{Naim2015, mandulak2024efficient}. The best linear time approximation algorithm is based on scaling techniques and achieved a $(1-\epsilon)$-approximation for arbitrarily small constant $\epsilon$~\cite{duan_linear-time_2014}. However, this complex algorithm is not shown to perform better than the 1/2-approximate ones for real world graph~\cite{al20222}.

\subsection{\textbf{Semi-streaming Matching:}}
\label{ssec:stream-match}
Motivated by solving extremely large matching problems, the most recent developments on matching includes semi-streaming computations, where the edges are streamed one by one and and an algorithm is allowed to use only $O(n \log n)$ space. The semi-streaming model for graph algorithms was proposed in the seminal work of Feigenbaum et al~\cite{feigenbaum2005graph}, where they developed a 1/6-approximation algorithm for matching. These were improved in subsequent work, culminating to the breakthrough 1/2-approximation algorithm of Paz and Schwartzman (Henceforth, PS algorithm)~\cite{PazS17}. The PS algorithm  can also be interpreted through primal-dual paradigm~\cite{GhaffariW19}, which provides a post-posterior instance-wise quality guarantee. Ferdous et al.~\cite{ferdousStreaming} compared the PS algorithm against state-of-the-art approximate matching algorithms and showed that it is extremely efficient in execution time and memory usage.

In this paper, we integrate three approximate matching algorithms (GD, LD, and PS) into \tokenjoin, a state-of-the-art set similarity join workflow, and demonstrate the effectiveness of approximate matching algorithms. To the best of our knowledge, this is the first such work for set similarity join.

\section{Details on Threshold Join. }\label{sec:methods-threshold}

The threshold based Fuzzy Join workflow can be divided into several components which we describe in this section. Our discussion is primarily based on the \tokenjoin~ method~\cite{TokenJoinOG}. Given a constant $\delta \in [0,1]$ that represent the admissibility of a pair into the join, we can convert an equivalent constant using the matching score $|R \:\widetilde{\cap}_{\phi}\:S|$ for a query set $R$ and a candidate set $S$ as described in \S\ref{ssec:set_sim_prelim}.

\subsection{\textbf{Candidate Generation:}} For candidate generation, an inverted list $I$ is built using $\mathcal{D}$ to map each token to the sets that contain that token, sorted in increasing order by size. Candidates are pulled from $I$ based on a size filter $|R|/\delta$. For each candidate $S$, if $S$ is already marked as a candidate, its utility score is increased; otherwise, $S$ is added to the set of candidates with its utility score initialized. The sum of possible utility scores is kept as $\sigma$ and is decremented for each candidate pairing's utility score added. 
Candidate generation stops when $\sigma$ falls below the threshold $\theta = \frac{2 \cdot \delta}{1 + \delta}$.

\subsection{\textbf{Candidate Refinement:}} Following generation, refinement seeks to apply the matching threshold $\theta_{RS} = \frac{\delta}{1+\delta}(|R| + |S|)$ to each candidate $S$. Based on the utility score set in candidate generation, $S$ can be directly pruned if its utility score and $\sigma$ fall below $\theta_{RS}$. This upper bound is then repeatedly refined per token, updating $\sigma$ and checking if a token $t \in S$. If so, its utility score is increased; otherwise, the upper bound is check and $S$ is pruned if its upper bound is below $\theta_{RS}$.

\subsection{\textbf{Additional Filtering:}}
The authors of \cite{TokenJoinOG} extend the refinement method with both a \textit{Positional Filter} and a \textit{Joint Utility Filter}, extending the \tokenjoin~ method to TJPJ. First, the positional filter relies on the idea that the token's position can be used in the inverted index $I$ to tighten the upper bound on a candidate. Thus, $I$ is adapted to store the position where the token was found as well as the token itself. The Joint Utility filter generalizes candidate pair utility values, that are normally set to a candidate $S$, to a pairing $(R,S)$, restricting the resultant matching by  $l = \text{min} (|R|,|S|)$ edges in the bipartite graph. Thus, this gives a better bound on the utility values, as the top $l$ values can be picked. These additions are included in the candidate refinement for better consideration of both the elements in $R$ and in $S$. 

\section{Memory and Parallelization Experiments}

\subsection{Memory Usage Analysis}\label{ssec:results-memory}
To analyze the memory usage patterns of the approximation and exact methods, we assess the average and peak memory consumption, and examine memory consumption throughout the program execution, in upcoming \S\ref{sssec:results-peak-memory} and \S\ref{sssec:results-dist-memory}.

\subsubsection{\textbf{Peak memory usage:}}\label{sssec:results-peak-memory}
\begin{figure}[!ht]
    \centering
    \includegraphics[scale=0.42]{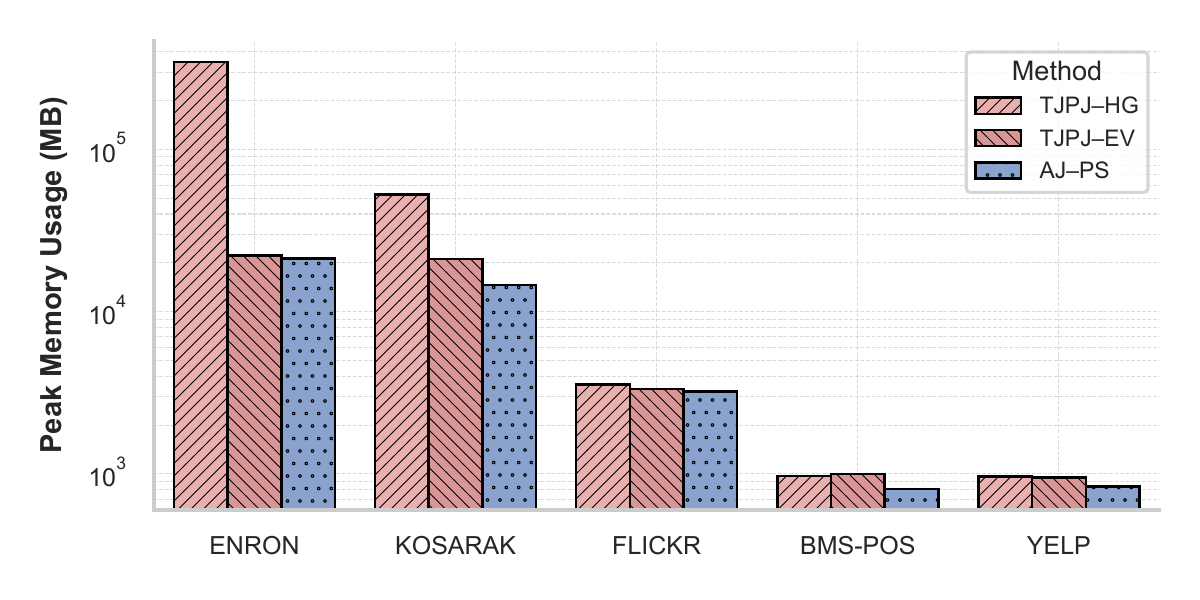}
    \caption{Comparison of memory usage using the integrated \ajps verification with the default \tjhg and \tjev cases at $|\mathcal{R}|=\text{20\%}$ (lower is better). Average memory usage reduction is 23\% in \ajps compared to \tjhg\slash \tjev.}
    \label{fig:memory-comparison}
\end{figure}

Given that \ajps does not require the entire graph being built in advance, we compare its peak memory usage relative to the exact \tjhg and \tjev methods. For peak memory analysis depicted in Fig.~\ref{fig:memory-comparison}, we run each method on a 20\% sample of a subset of datasets using JAC. We use the built in \texttt{tracemalloc} snapshots\linkfoot{https://docs.python.org/3/library/tracemalloc.html} to obtain peak memory usages. Notably, we observe the highest improvement compared to \tjhg for the denser instances, with up to a 93\% and 72\% reduction in memory usage on ENRON and KOSARAK, respectively. On average, \ajps yields a 34\% reduction compared to \tjhg and a 12\% reduction against \tjev, improving the memory usage for every instance considered. 

\subsubsection{\textbf{Memory usage across program lifetime:}}\label{sssec:results-dist-memory}
\begin{figure}[!ht]
    \centering
    \includegraphics[scale=0.55]{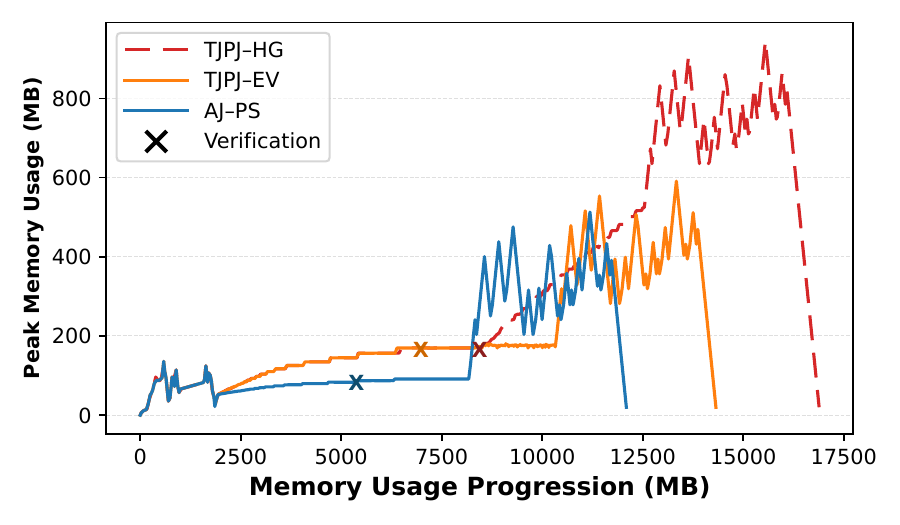}
    \caption{Memory usage progression relative to peak memory usage during a run of \tjhg\slash \tjev and \ajps using KOSARAK $|\mathcal{R}|=\text{10\%}$. Verification center is marked by ``$\times$''.}
    \label{fig:memory_dist}
\end{figure}

In Fig.~\ref{fig:memory_dist}, we compare the lifetime memory consumption (from beginning to end of a particular program run) of the exact and approximate methods on an instance of $|\mathcal{R}|=\text{10\%}$ for KOSARAK (one of the denser instances, see Table~\ref{tab:data}). We collect up to 500 distinct samples of instantaneous memory usage for \tjhg, \tjev and \ajps using the \texttt{valgrind} \texttt{Massif} heap profiling tool\linkfoot{https://valgrind.org/docs/manual/ms-manual.html}. 
Verification happens in the middle portion of the computation (denoted by ``x'' in Fig.~\ref{fig:memory_dist}), where \ajps shows under 100MB of peak memory usage while \tjhg\slash \tjev can consume up to 40\% extra memory in this phase.  For KOSARAK (best performing dataset for the approximate methods),  \tjhg, \tjev and \ajps depicts peak memory usage of 944MB, 591MB and  512MB, respectively; \ajps exhibits 14\slash 45\% better memory footprint.

\subsection{Preliminary parallelization of \ajps}\label{ssec:results-parallel}

Designing highly concurrent approximation algorithms is an active research area~\cite{pothen2019approximation}. As such, there is parallelism potential in the approximate matching methods used in verification. Since \ajps is the best performing algorithm in this scenario (both in terms of performance and memory), we develop a ``proxy'' verification pipeline comprising of a preliminary parallel version of \ajps using OpenMP thread-parallelism~\cite{dagum1998openmp}, leveraging randomly generated data to analyze the parallel efficiency across varied data ranges. This ''proxy'' pipeline simulates the graph building (i.e., edge streaming, in the context of \ajps) and matching phases of verification on a set of 500 randomly generated pairings of $\{R,S\}$. Within these randomly generated pairings, we vary the elements\slash set between $10^2$--$10^6$, imitating diverse dataset densities prevalent on real world datasets.

In terms of parallel implementation, we augment the streaming process of Lines \ref{alglinePSV:R_for_loop} to \ref{alglinePSV:stream_process} in Algorithm \ref{alg:ps_verification} by performing set similarity calculations (JAC) in parallel and storing the edge similarity information in a buffer. This buffer acts as our stream, where we have a single thread pull the edges sequentially and commit to the stack. Given the potential for hazards while accessing the buffer in multithreaded environments, we implement thread locks for task synchronization.
\begin{figure}[!ht]
    \centering
    \includegraphics[scale=0.60]{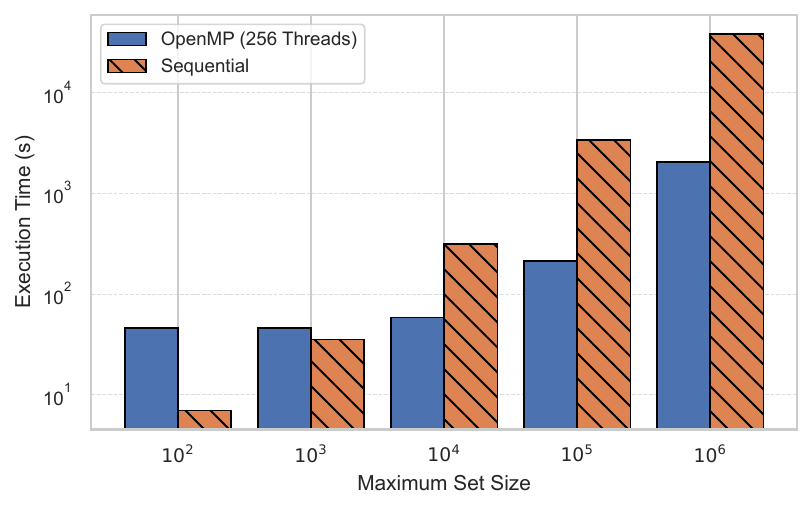}
    \caption{Preliminary execution time results (lower is better) of the proxy OpenMP-based \ajps verification (on 256 threads) vs. the sequential, varying the set density. }
    \label{fig:proxy_res}
\end{figure}
As shown in Fig.~\ref{fig:proxy_res}, for smaller instances (e.g., set sizes less than 1K), threads are starved (of work), and parallelism overheads outweigh any potential benefits. However, when the set sizes increase, parallel efficiency improves, enhancing the performance by 8$\times$ on average as compared to the default serial version on 256 threads (observed maximum speedup of 18$\times$ at set size $10^6$). Using a single thread for stream update and subsequent lock-based synchronization in this version limits ideal scalability, which could be mitigated by an enhanced parallel implementation, maintaining a stack per thread\slash vertex, at the expense of extra memory.


\begin{figure}
    \centering
    \includegraphics[scale=0.35]{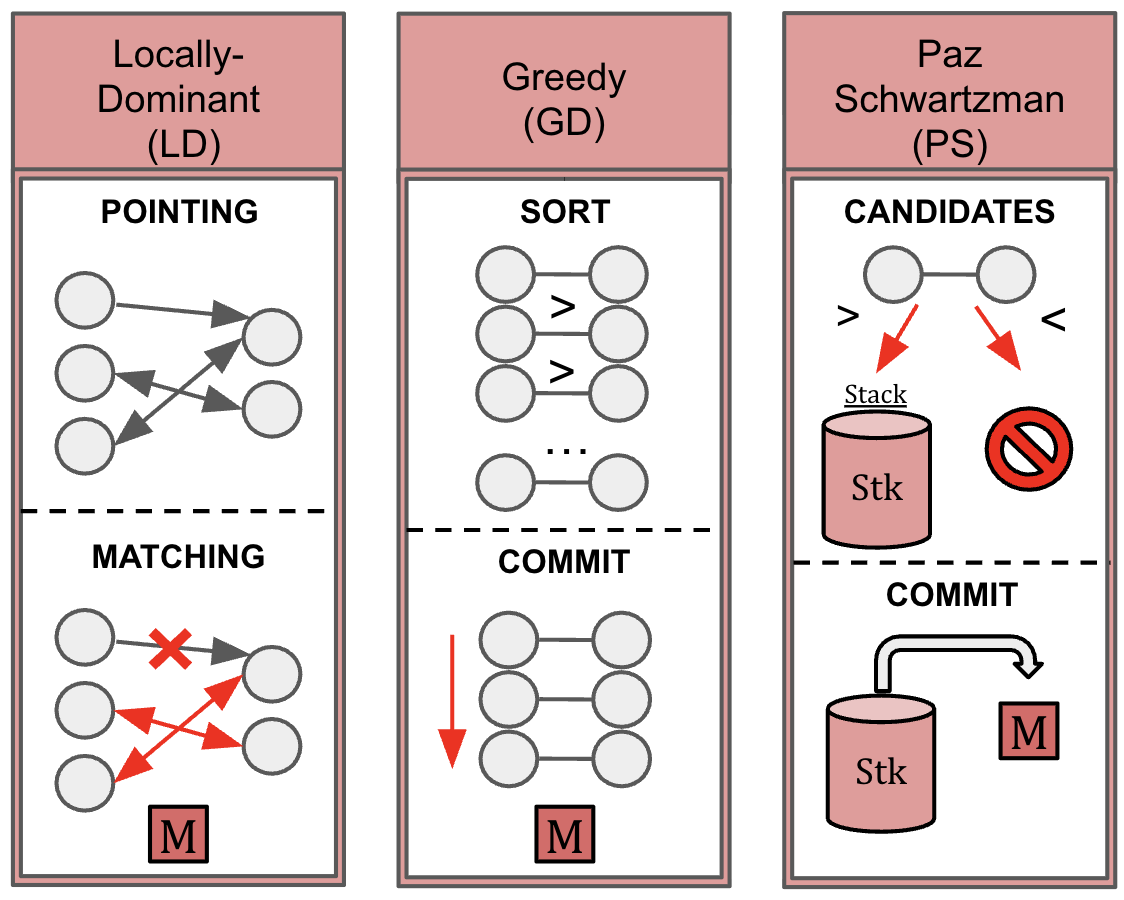}
    \caption{Generalized matching process abstraction for each approximate method, $M$ is the matching.}
    \label{fig:verifyCartoon}
\end{figure}

\section{Omitted Pseudocodes}

\begin{algorithm}[t]
\caption{Set Verification}\label{alg:verify}
\begin{algorithmic}[1]
\AlgIn{Candidate Sets: $R,S$, Threshold: $\theta_R$}
\AlgOut{Score: $sim_\phi(R,S)$}
\State ov $= |R \cap S|$ \Comment{Phase 1: Deduplication} \label{alglineV:overlap}
\State $R_d, S_d = \texttt{deduplication}(R,S)$
\If {$R_d = \emptyset$}
    \State \Return $\frac{\text{ov}}{(|R| + |S| - \text{ov})}$ \label{alglineV:dedupReturn}
\EndIf
\State $G(V,E,w) = \emptyset, \text{UB} = |R|$ \Comment{Phase 2: Graph Building}
\label{alglineV:graphBuild}
\ForAll {$r \in R_d$}
\label{alglineV:R_for_loop}
    \State $max_s = 0$
    \ForAll{$s \in S_d$}
        \State $score = \texttt{sim}(r,s)$
        \State $max_s = \texttt{max}(max_s,score)$
        \State $V= V \cup r \cup s$ 
        \State $E = E \cup \{r,s,score\}$ 
        \label{alglineV:addEdge}\Comment{Add edge to graph}
    \EndFor
    \State $\text{UB} = \text{UB} - (1-max_s)$
    \If{$\theta_R > \text{UB}$}
        \State \Return $\frac{\text{UB}}{(|R| + |S| - \text{UB})}$ \label{alglineV:ubReturn}
    \EndIf
\EndFor
\State $W_M = \text{ov} + \texttt{matching}(G)$ \Comment{Phase 3: Bipartite Matching} \label{alglineV:matching}
\State \Return $\frac{W_M}{(|R| + |S| - W_M)}$

\end{algorithmic}
\end{algorithm}

\begin{algorithm}[!ht]
    \caption{Locally-Dominant (LD) Matching}
    \label{alg:ld}
    \begin{algorithmic}[1]
    \AlgIn{Graph: $G(V,E,w)$}
    \AlgOut{A locally dominant matching in $mate$ array}
        \State $M \gets \emptyset$
        \While{$G$ is not empty}
        
        \ForAll {$v \in V$}\Comment{Phase 1: Pointing}
            \State $\mathit{mate}(v)=\argmax_{u \in \calN(v)} w(\{u,v\})$ 
        \EndFor
        
        \ForAll {$e(u,v) \in E$}\Comment{Phase 2: Matching}
            \If {$\mathit{mate}(v) = u$ and $\mathit{mate}(u) = v$} \Comment{LD edge} \label{line:ld}
                \State $M = M \cup e$
                \State $G = G \setminus \{ e \cup \calN(e) \}$
            \EndIf
        \EndFor
        \EndWhile
    \end{algorithmic}
\end{algorithm}
\end{document}